\documentclass[10pt]{article}

\usepackage[utf8]{inputenc}
\usepackage{graphicx}
\usepackage{geometry} %Enable changing the margins of the page
\usepackage{amsmath} %Enables equation*
\usepackage{amssymb} %Enables expectation operator
\usepackage{commath} %Enables abs
\usepackage{float} %Enables H for strict float (table, graph) location
\usepackage{multirow} %Enables multirow cell in a table (i.e. group multiple rows of cells)
\usepackage{parskip} %Enables line skip at the end of a paragraph
\usepackage[table]{xcolor} %Enables alternating row tables
\usepackage{courier} %Enbles \texttt{} for courier font
\usepackage{subcaption} %Enables subfigures
\usepackage{hyperref} %Enables hyperlinks
\usepackage{listings} %Used to input code
\usepackage{todonotes} %For TODOs
\usepackage{biblatex} %For citing
\usepackage{lipsum} %For random text
\usepackage{titlesec} %To change the spacing between titles
\usepackage{multicol}

\titlespacing*{\section}{0pt}{0.1\baselineskip}{0.1\baselineskip}
\titlespacing*{\subsection}{0pt}{0\baselineskip}{0\baselineskip}

\newgeometry{vmargin={1.5cm}, hmargin={1.5cm}}
\renewcommand\lstlistingname{Figure}
\begin{document}
\definecolor{comment}{RGB}{0,128,0} % dark green
\definecolor{string}{RGB}{255,0,0}  % red
\definecolor{keyword}{RGB}{0,0,255} % blue

\lstdefinestyle{c}{
	commentstyle=\color{comment},
	stringstyle=\color{string},
	keywordstyle=\color{keyword},
	basicstyle=\footnotesize\ttfamily,
	numbers=left,
	numberstyle=\tiny,
	numbersep=5pt,
	frame=lines,
	breaklines=true,
	prebreak=\raisebox{0ex}[0ex][0ex]{\ensuremath{\hookleftarrow}},
	showstringspaces=false,
	upquote=true,
	tabsize=2,
}

\definecolor{mygreen}{RGB}{28,172,0} % color values Red, Green, Blue
\definecolor{mylilas}{RGB}{170,55,241}

\lstdefinestyle{matlab}{
    %basicstyle=\color{red},
    breaklines=true,%
    morekeywords={matlab2tikz},
    keywordstyle=\color{blue},%
    morekeywords=[2]{1}, keywordstyle=[2]{\color{black}},
    identifierstyle=\color{black},%
    stringstyle=\color{mylilas},
    commentstyle=\color{mygreen},%
    showstringspaces=false,%without this there will be a symbol in the places where there is a space
    frame=lines,
    numbers=left,%
    numberstyle={\tiny \color{black}},% size of the numbers
    numbersep=9pt, % this defines how far the numbers are from the text
    emph=[1]{for,end,break},emphstyle=[1]\color{red}, %some words to emphasise
    %emph=[2]{word1,word2}, emphstyle=[2]{style},    
}
\begin{multicols}{2}[
 %Add title and authors
 {\centering
 \LARGE \textbf{Real-Time Speech Enhancement Using Spectral Subtraction with Minimum Statistics and Spectral Floor}\\[0.4cm]
 \large Georgios Ioannides$^{1}$ georgios.ioannides16@alumni.imperial.ac.uk and Vasilios Rallis\footnote[1]{Equal Contribution} vasilios.rallis98@gmail.com\\[0.1cm]
% \large Date: \today\\[0.4cm]
 }
]

%Replace listings counter with figure counter this must be done after the /begin{document}
\renewcommand{\thelstlisting}{\thefigure}

\section{Abstract}
An initial real-time speech enhancement method is presented to reduce the effects of additive noise. The method operates in the frequency domain and is a form of spectral subtraction. Initially, minimum statistics are used to generate an estimate of the noise signal in the frequency domain. The use of minimum statistics avoids the need for a voice activity detector (VAD) which has proven to be challenging to create \cite{martin1994spectral}. As minimum statistics are used, the noise signal estimate must be multiplied by a scaling factor before subtraction from the noise corrupted speech signal can take place. A spectral floor is applied to the difference to suppress the effects of "musical noise" \cite{boll1979suppression}. Finally, a series of further enhancements are considered to reduce the effects of residual noise even further. These methods are compared using time-frequency plots to create the final speech enhancement design.

\section{Introduction}
Background additive noise that has distorted a speech signal can degrade the performance of many real-world digital communication systems. Today, digital communication systems are increasingly being used in noise environments such as vehicles, factories and airports. Signal Processing techniques are also used in brain modelling applications\cite{spatio}. Robustness to noise sensitivity have become key properties in any communication system. In this work, a real-time spectral subtraction system will be implemented to reduce the background noise in a speech signal while leaving the speech itself intact. This is known as speech enhancement.

\section{Basic Implementation}
\subsection{High-level Overview}
In spectral subtraction, the assumption is that the speech signal $s(t)$ has been distorted by a noise signal $n(t)$ with their sum being denoted by $x(t)$ (\ref{eq:3.1.1}).

\begin{equation}
    x(t) = s(t) + n(t)\label{eq:3.1.1}
\end{equation}

In the frequency domain, these signals become:

\begin{equation}
    X(\omega) = S(\omega) + N(\omega)\label{eq:3.1.2}
\end{equation}

where $X(\omega)$, $S(\omega)$ and $N(\omega)$ are the Fourier transforms of $x(t)$, $s(t)$ and $n(t)$ respectively. Effectively, the spectral subtraction method operates in the Fourier domain by attempting to subtract an estimate of $N(\omega)$ which will be denoted as $\hat{N}(\omega)$ from $X(\omega)$, to produce a final signal $Y(\omega)$ (\ref{eq:3.1.3}).

\begin{equation}
    Y(\omega) = X(\omega) - \hat{N}(\omega)\label{eq:3.1.3}
\end{equation}

However, since the phase of the noise is not known, only the magnitude of the noise estimate $\hat{N}(\omega)$ will be subtracted from $X(\omega)$ leaving the phase of $X(\omega)$ distorted by the noise (\ref{eq:3.1.4}).

\begin{align}
    Y(\omega) &= X(\omega) - \abs{\hat{N}(\omega)}\nonumber\\
              &= X(\omega)\left(1 - \frac{\abs{\hat{N}(\omega)}}{\abs{X(\omega)}}\right)\nonumber\\
              &= X(\omega)g(\omega)\label{eq:3.1.4}
\end{align}

An issue with simply implementing (\ref{eq:3.1.4}) is that if $g(\omega)$ is negative for some frequency bins, the phase of those frequency bins will be shifted by $\frac{\pi}{2}$ radians. As stated by \cite{hartmann2004signals}, the relative phases of two signal components is relevant if the two components are separated by less than a critical bandwidth. This critical bandwidth is close to $1/6^{\text{th}}$ of an octave after 1kHz. Therefore, under some conditions, phase distortion might result in an audible distortion in the time domain. A solution to this problem would be to modify $g(\omega)$ to:

\begin{equation}
    g(\omega) = \text{max}\left(0,1 - \frac{\abs{\hat{N}(\omega)}}{\abs{X(\omega)}}\right)\label{eq:3.1.5}
\end{equation}

\begin{figure}[H]
\centering
\includegraphics[width=1\linewidth]{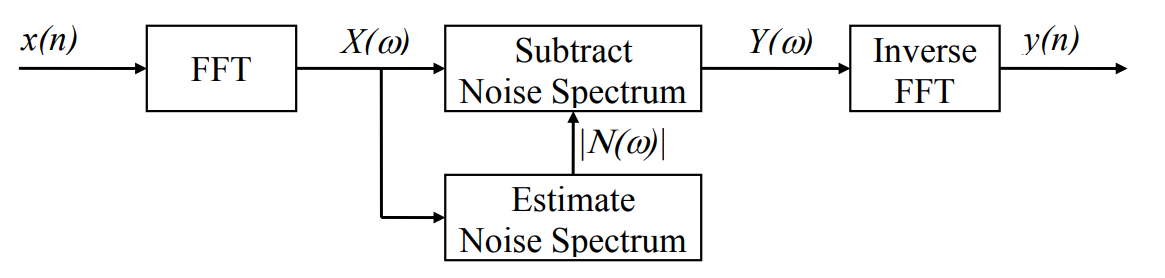}
\caption{Block diagram of spectral subtraction \cite{mitchProjNotes}}
\label{fig:3.1.1}
\end{figure}

Throughout this work, $g(\omega)$ will be modified in search of an improvement in intelligibility of of the final signal $y(t)$. Ideally, $Y(\omega) \approx S(\omega)$, thus, using the inverse Fourier transform, the original speech signal $s(t)$ can be recovered. The process is illustrated in Figure \ref{fig:3.1.1}

The assumption of additive noise implies that $n(t)$ and $s(t)$ are statistically independent \cite{princOfDigCom}. This assumption can be applied in most real-world situations as no knowledge of the probability density function (PDF) or the frequency domain of the noise is required.

\subsection{Frame Processing}
For the system to be real-time, the speech signal $s(t)$ must first be split into smaller sections so that the processing can take place before the entire signal has arrived. These smaller sections are called frames and their size is denoted as $N$. For the basic implementation, $N=256$. It is critical that $N$ is a power of 2 so that the radix-2 FFT algorithms can be used. This leads to a reduction in the time-complexity of the FFT algorithm from $O(N^2)$ to $O(N\log(N))$. The reduction in the run-time of the algorithm for large values of $N$ (i.e. $N>100$) is the critical for the system to be achievable in real-time.

\begin{figure}[H]
\centering
\includegraphics[width=1\linewidth]{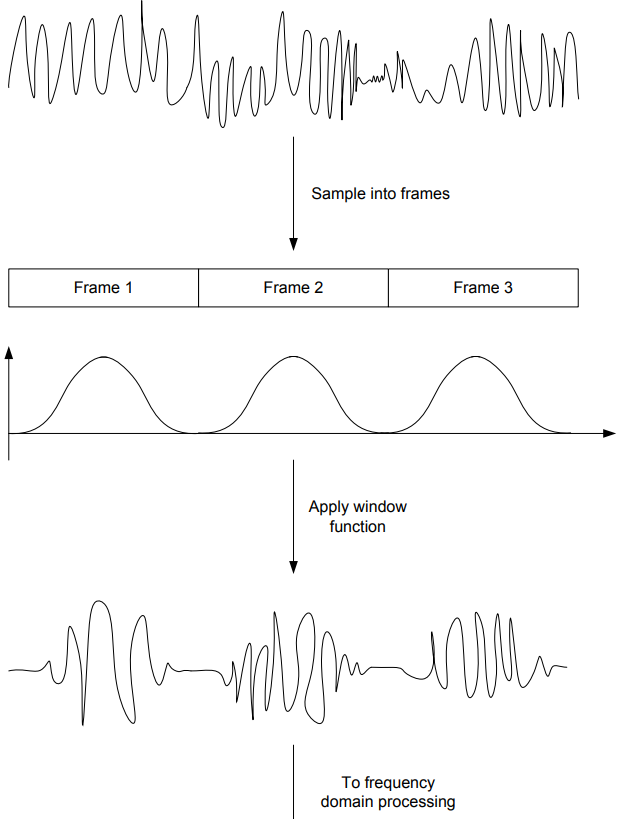}
\caption{Problem with simply applying window in the time domain \cite{mitchProjNotes}}
\label{fig:3.2.1}
\end{figure}

However, the discontinuities at the edges of each frame will lead to spectral artifacts. To solve this issue, a window is applied in the time domain before the FFT of the frame is computed. Nevertheless, by windowing $x(t)$ in the time domain, the signal has been distorted; thus, as shown in Figure \ref{fig:3.2.1}, the original time domain signal will not be recovered.

\begin{figure}[H]
\centering
\includegraphics[width=1\linewidth]{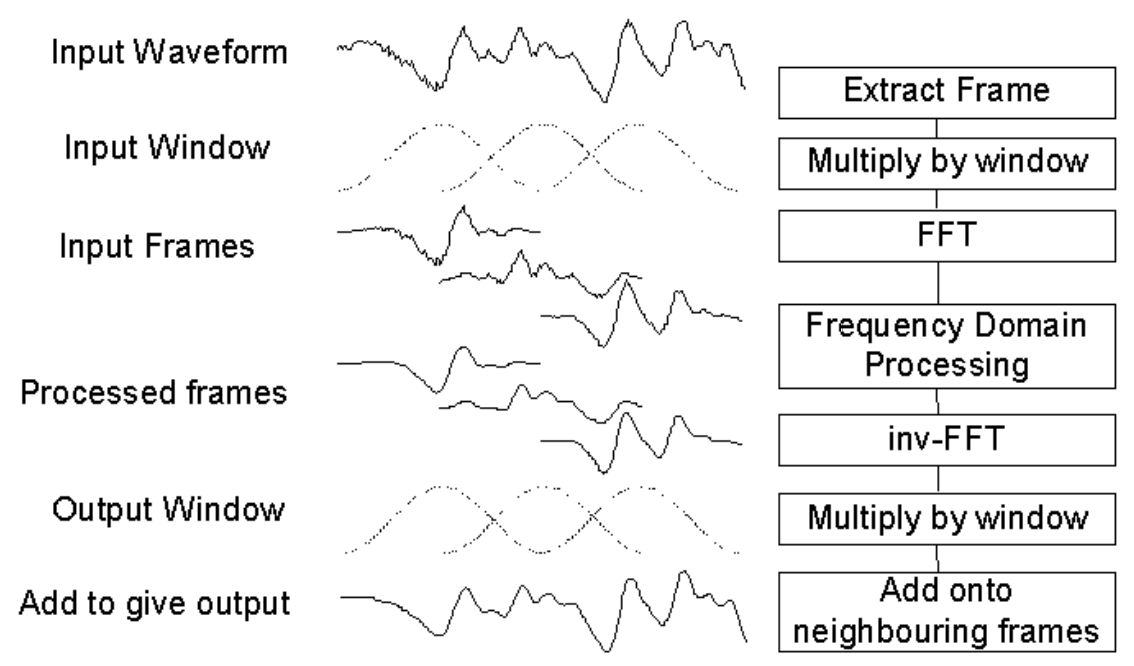}
\caption{Overlap and add process \cite{mitchProjNotes}}
\label{fig:3.2.2}
\end{figure}

To solve this, the individual frames can be overlapped so that the sum of overlapping windows is always 1. The number of frames that overlap is known as the oversampling factor. The process for an oversampling factor of 2 is shown in Figure \ref{fig:3.2.2}. Note that for the basic implementation of the spectral subtraction algorithm, an oversampling factor of 4 was used instead (i.e. each frame will contain $256/4=64$ new samples)

\begin{figure}[H]
\centering
\includegraphics[width=1\linewidth]{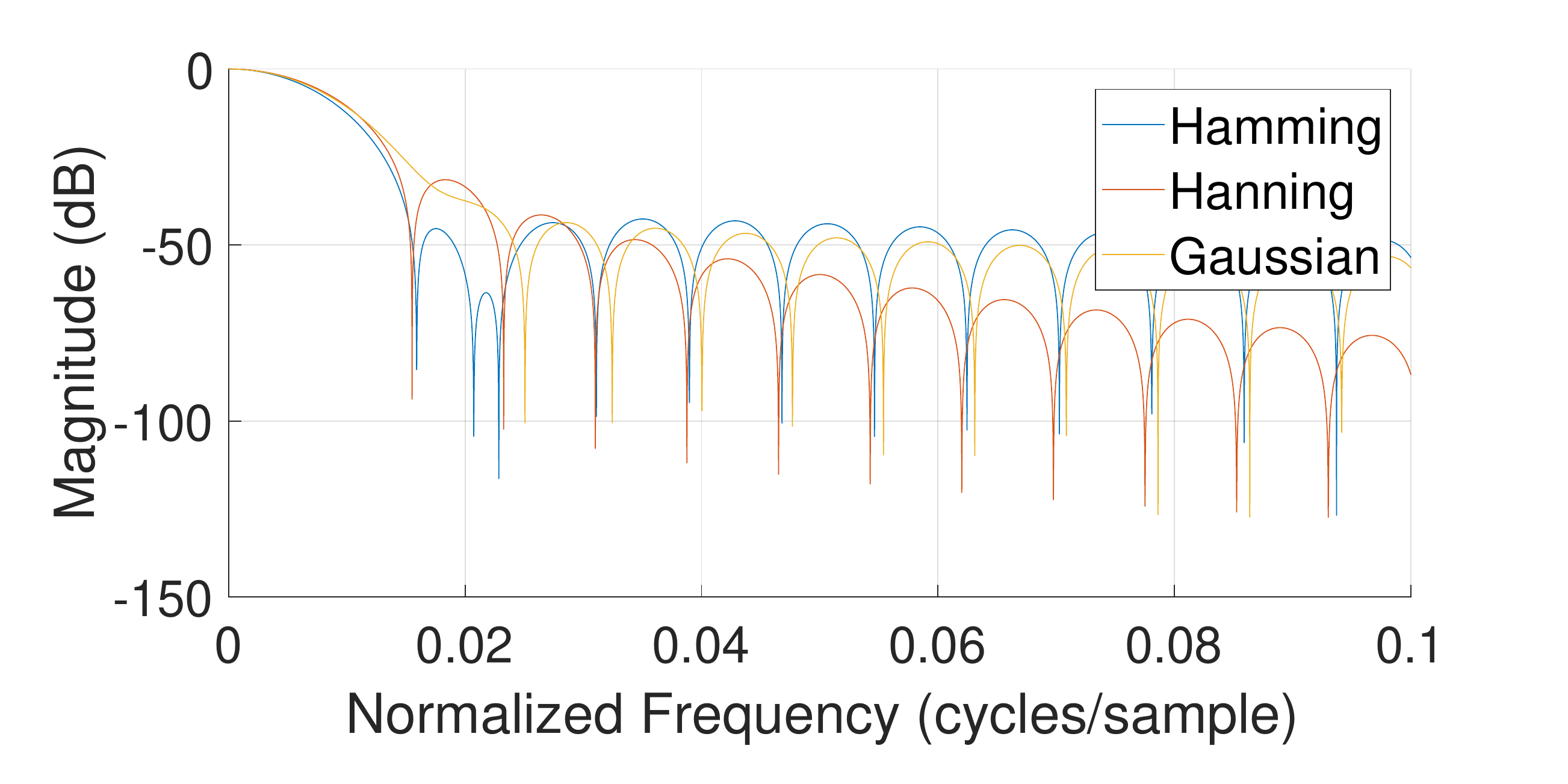}
\caption{Frequency domain of Hamming, Hanning and Gaussian Windows}
\label{fig:3.2.3}
\end{figure}

As shown in Figure \ref{fig:3.2.2}, another time domain window is applied to $x(t)$ after the Inverse Fast Fourier Transform (IFFT) of the function is taken. This second window is necessary as a modification in the frequency domain is equivalent to filtering in the time domain which might lead to discontinuities when the frames meet. In implementations developed in this work, for both the input and output windows, the square root of the Hamming window was used (\ref{eq:3.2.1}). The Hamming window offers a relative first sidelobe amplitude level of -40.0dB Figure \ref{fig:3.2.3}. 

\begin{equation}
    \small
    w(t) = \sqrt{1-0.85185\cos\left(\frac{(2t+1)\pi}{N}\right)} \text{ for } t = 0,...,N-1\label{eq:3.2.1}
\end{equation}

\subsection{Noise Estimation}
As mentioned previously, for spectral subtraction to be performed, an estimation $\hat{N}(\omega)$ of the noise present in the signal is required (\ref{eq:3.3.1}). One way of finding this estimate would be to use a Voice Activity Detector (VAD) which detects whether speech is present in the signal and then take the average of all the frames where speech is not present \cite{boll1979suppression}. However, spectral subtraction based on VAD is exceptionally difficult to make so an easier approach is chosen \cite{martin1994spectral}.

For each frequency bin of $X(\omega)$, the minimum magnitude over the last 10 seconds is determined. This frame will be referred to as the Minimum Magnitude Spectral Estimate (MMSE) henceforth. Assuming that the speaker who is being recorded will make a brief pause within these 10 seconds to take a breath, the MMSE will correspond to the minimum magnitude of the realization of the noise within the speech pauses in the last 10 seconds. As this estimator will use the minimum of the noise realizations, it will severely underestimate the average magnitude of the noise signal. For this reason, a compensating factor denoted by $\alpha$ must be introduced. Since $x(t)$ is sampled at $8$kHz, and each new frame contains 64 new samples (i.e. 8ms of new information), 1250 frames must be stored in memory to find the MMSE. This is infeasible due to hardware limitations of the system in use.

A simplification can be made by storing just 4 frames denoted as $M_i(\omega)$ where $i = 1,..,4$. For each frame, $M_1(\omega)$ is updated by:

\begin{equation}
    M_1(\omega) = \text{min}\left(\abs{X(\omega)},M_1(\omega)\right)\label{eq:3.3.1}
\end{equation}

After 2.5 seconds (i.e. approximately 312 new frames), the frames are shifted and the new $M_i(\omega)$ takes the values of the previous $M_{i-1}$, for $i=4,...,2$ while $M_1(\omega)$ is set to $\abs{X(\omega)}$. The disadvantage of using this simplification is that the MMSE memory (i.e. how far into the past the minimum frequency bins will be searched for) will not be a constant 10 seconds since once the shift occurs, the new MMSE will have an effective memory of 7.508 seconds which will grow until it reaches 10 seconds and then reset again. Nevertheless, this is a small compromise for such a dramatic decrease in the amount of memory required.

\subsection{The noise trade-off}
As explained in \cite{berouti1979enhancement}, one of the problems with the implementation described above is the introduction of a new type of noise into $Y(\omega)$. This new type of noise will be referred to as musical noise. To explain this new type of noise, it is crucial to understand that there are peaks and valleys in the short-term power spectrum of the noise. Both the frequency and amplitude of these peaks will vary randomly from frame to frame. When spectral subtraction takes place according to $g(\omega)$ (\ref{eq:3.1.5}), depending on the value of $\alpha$ more peaks or more valleys will remain in the magnitude of the processed frame $\abs{X(\omega)}$. The peaks will be perceived as tones at a specific frequency. This frequency will change every frame, thus, for the implementation described above, the frequency of the tones will change every 8ms. The valleys will be perceived as broadband noise.

\begin{figure}[H]
\centering
\includegraphics[width=1\linewidth]{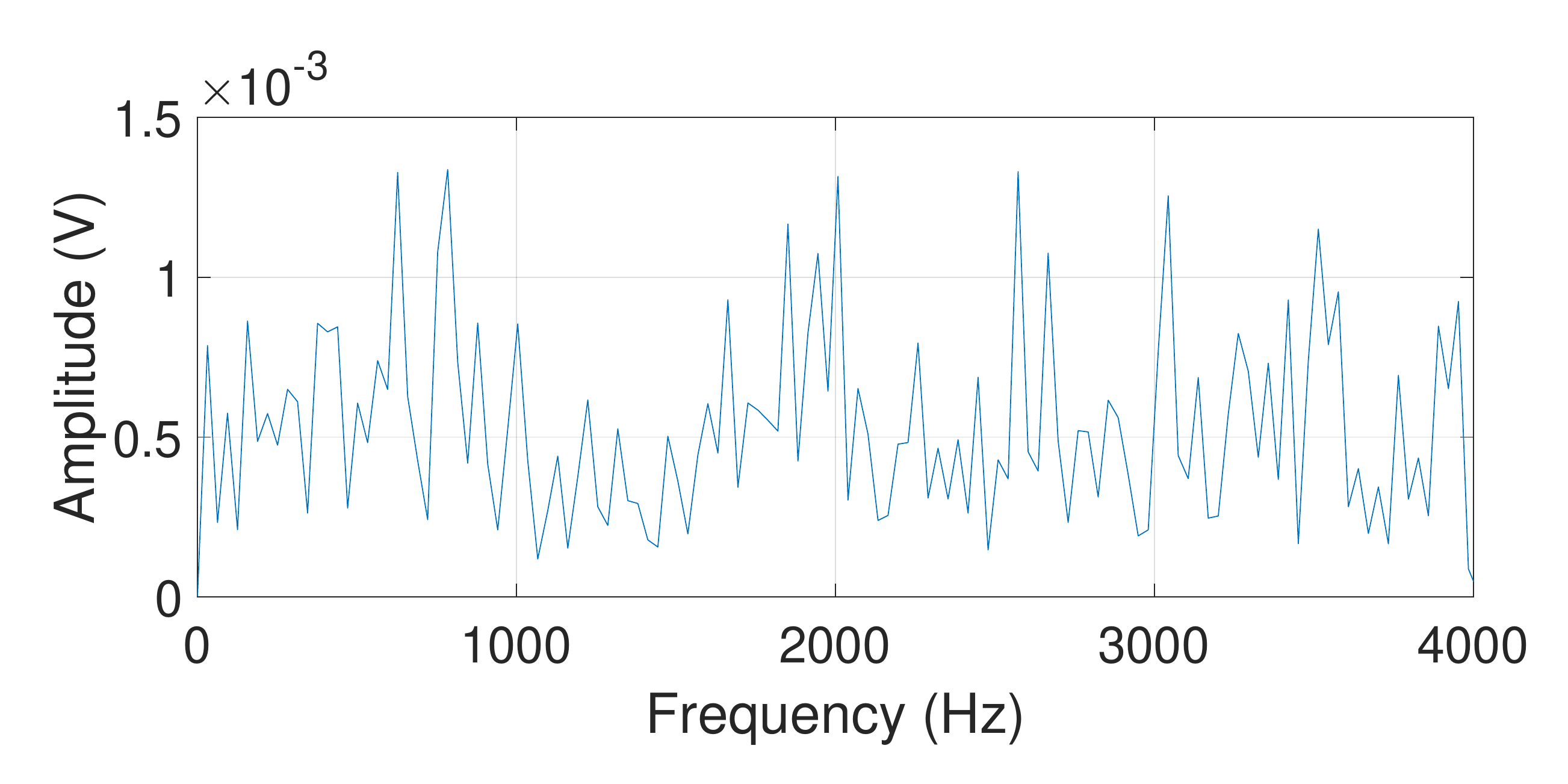}
\caption{MMSE over the past 1000 frames}
\label{fig:3.4.1}
\end{figure}

A simulated example is described to gain a better understanding. Using the MATLAB function \texttt{randn}, 1000 frame realizations of length 256 are generated. The MMSE over these 1000 frames is plotted in Figure \ref{fig:3.4.1}.

\begin{figure}[H]
\centering
\includegraphics[width=1\linewidth]{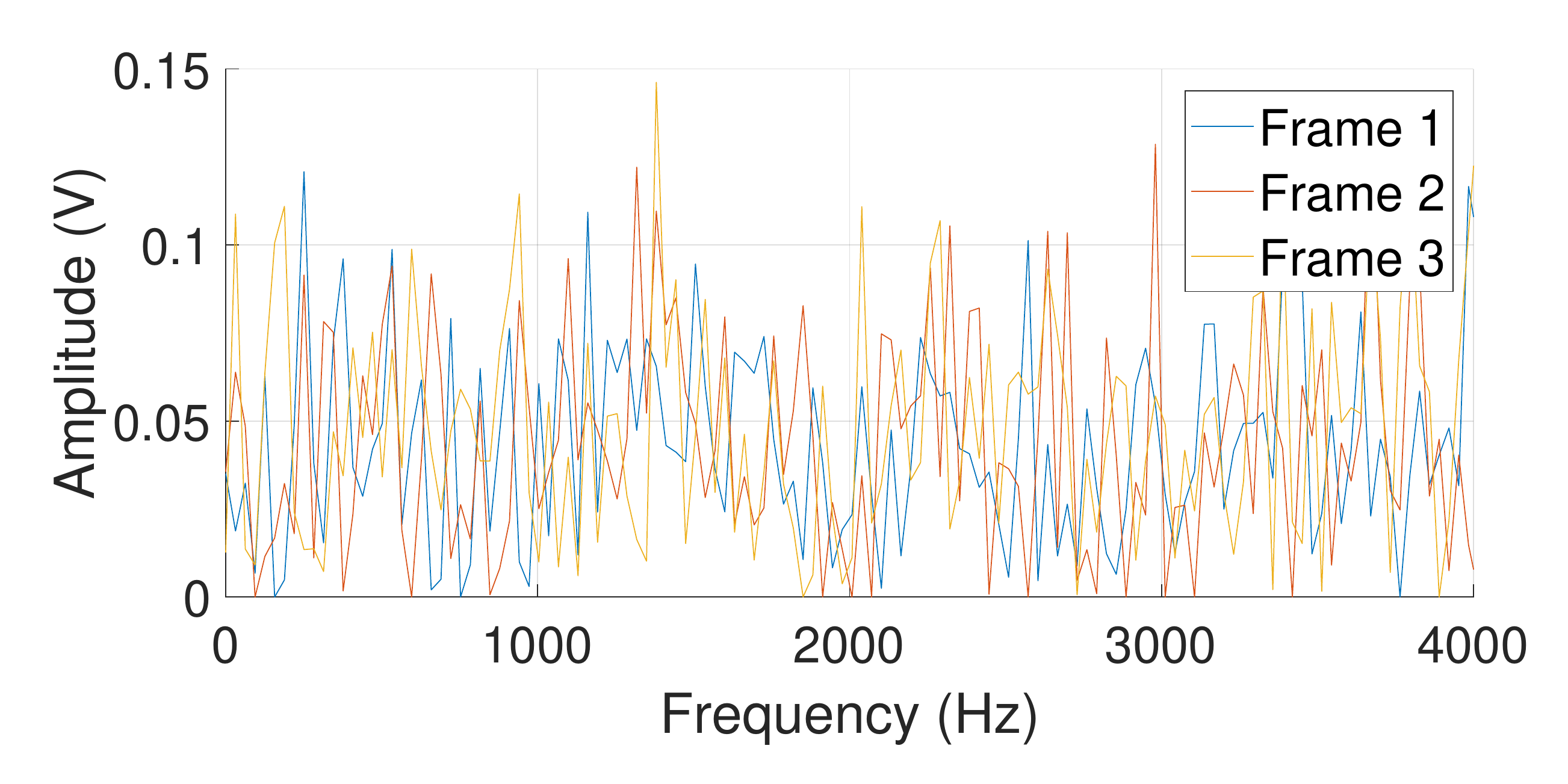}
\caption{Magnitude of three consecutive processed frames for $\alpha=20$}
\label{fig:3.4.2}
\end{figure}

The magnitude $\abs{Y(\omega)}$ of three consecutive processed frames for $\alpha = 20$ is plotted in Figure \ref{fig:3.4.2}. Note that both peaks and valleys are present in all three frames.

\begin{figure}[H]
\centering
\includegraphics[width=1\linewidth]{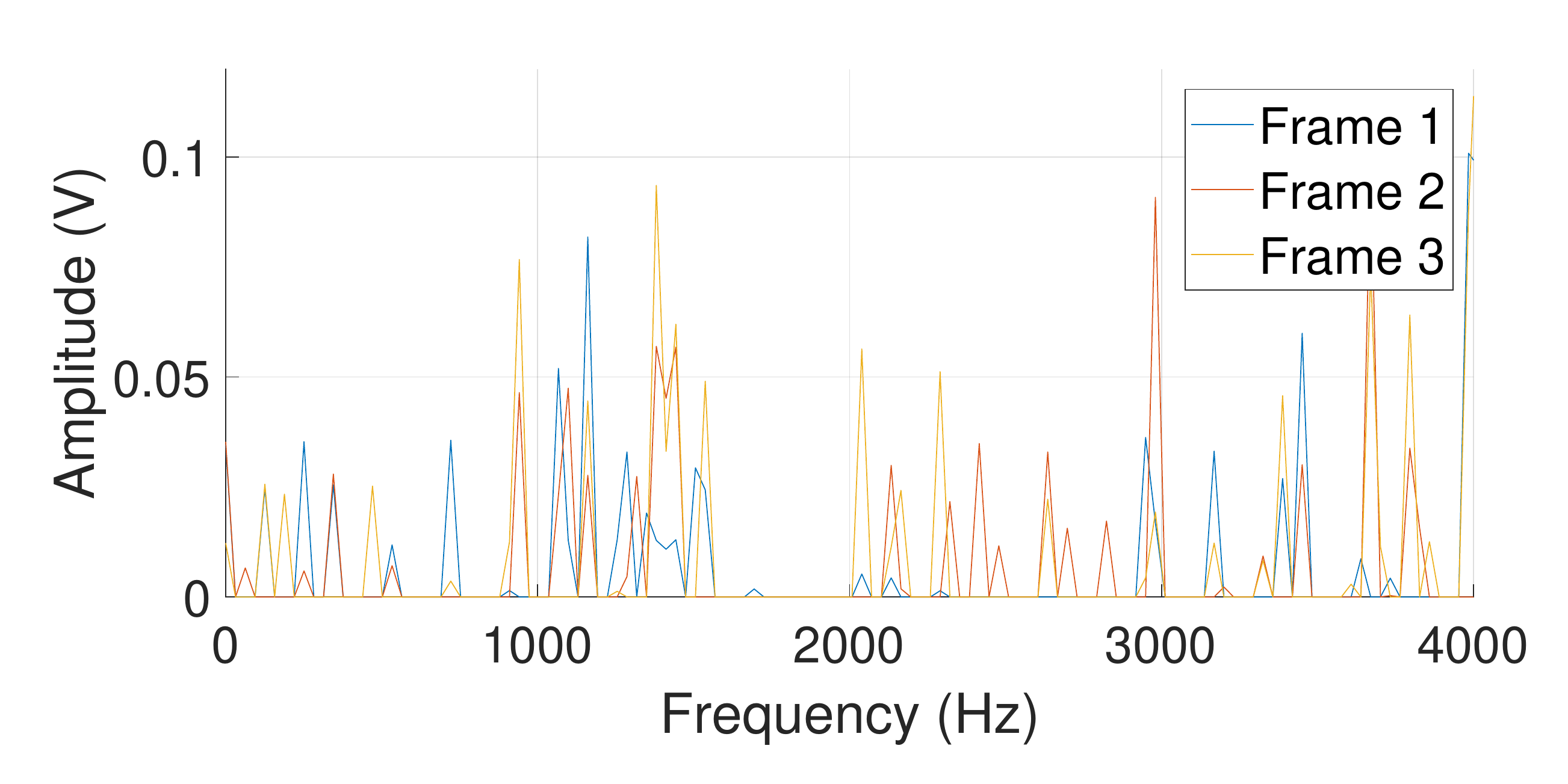}
\caption{Magnitude of three consecutive processed frames for $\alpha=200$}
\label{fig:3.4.3}
\end{figure}

By increasing the value of $\alpha$, the broadband noise in the frame will be suppressed while the effect of the musical noise (i.e. the peaks) will be further enhanced since it will not be masked by the broadband noise. The magnitude $\abs{Y(\omega)}$ of three consecutive processed frames for $\alpha=200$ is plotted in Figure \ref{fig:3.4.3}. As expected, the peaks are more prevalent even though their amplitude has been decreased.

A solution to this musical noise problem, is to further modify $g(\omega)$ to introduce a new parameter $\lambda$ which will be referred to as the spectral floor (\ref{eq:3.4.1}). Effectively, the parameter will be used to mask the musical noise with broadband noise (Figure \ref{fig:3.4.3}).

\begin{equation}
    g(\omega) = \text{max}\left(\lambda,1 - \alpha\frac{\abs{\hat{N}(\omega)}}{\abs{X(\omega)}}\right)\label{eq:3.4.1}
\end{equation}

\begin{figure}[H]
\centering
\includegraphics[width=1\linewidth]{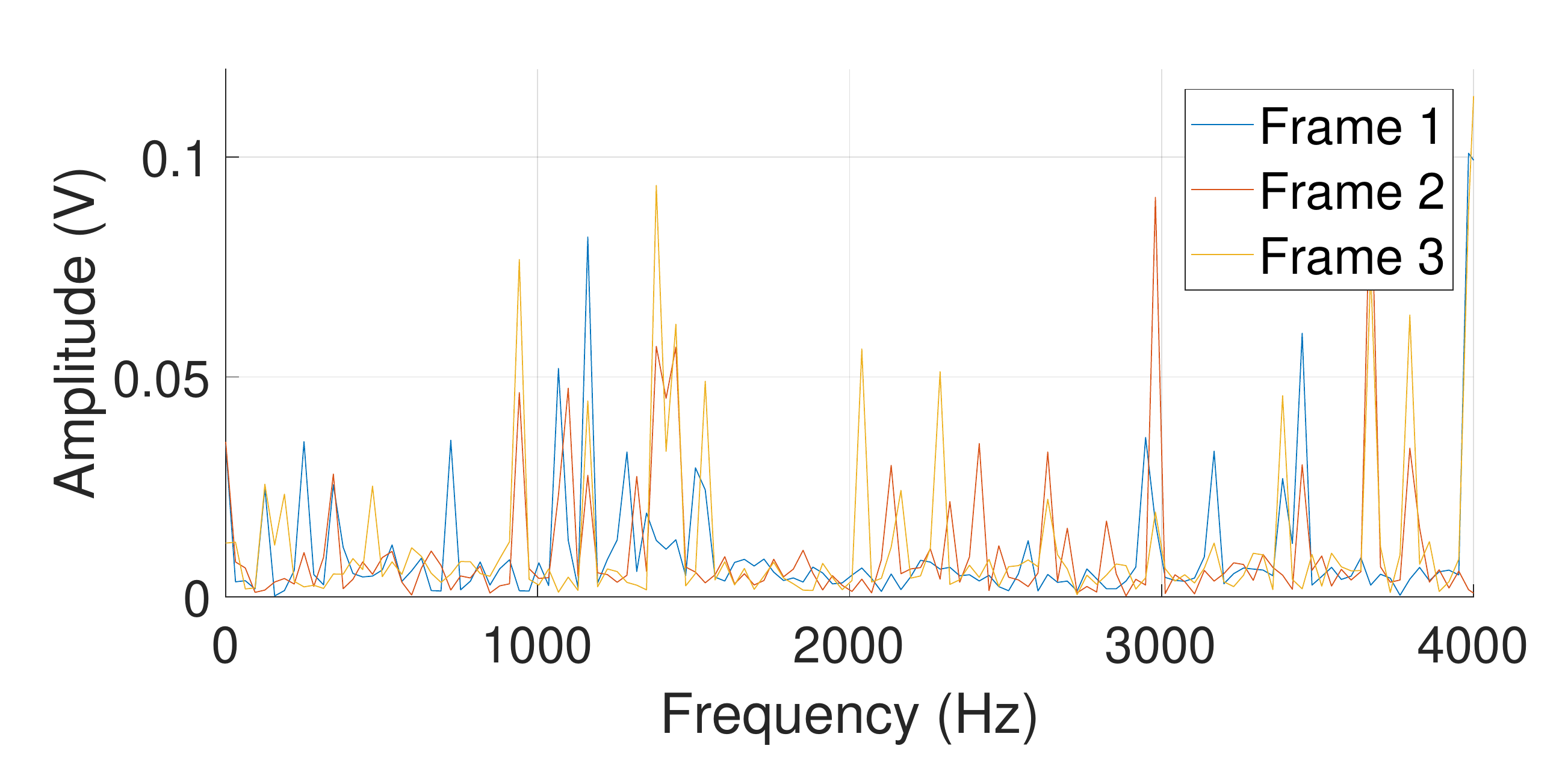}
\caption{Magnitude of three consecutive processed frames for $\alpha=200$ and $\lambda=0.1$}
\label{fig:3.4.4}
\end{figure}

Since the MMSE is used as an estimate for the noise, the appropriate value (i.e. the one that leads to best intelligibility of the speech) of $\alpha$ will increase with:

\begin{enumerate}
    \item The memory of the MMSE
    \item The variance of the noise. This equivalent to the power of the zero mean noise.
\end{enumerate}

In this simulated example, the signal consisted of only noise; however, it must be underscored that if $\alpha$ is too large, distortion caused by the spectral subtraction will decrease the speech intelligibility.

Overall, through the above analysis, it is clear that the parameters of spectral subtraction method used, must be adjusted to achieve a balance between, musical noise, broadband noise and speech intelligibility. This intuition will be used in the next sections to further improve the current implementation.

\subsection{Implementation in C}
The key parts of the C code for the basic implementation are described in the following section. The frame that must be processed is located in the \texttt{inframe} array. The first step is to move the frame from the \texttt{inframe} array to the \texttt{intermediate} array and convert the elements of \texttt{inframe} from \texttt{float} to \texttt{complex} which is a \texttt{struct} defined in the \texttt{complx.h} header file. The conversion from \texttt{float} to \texttt{complex} is required as the signature of the \texttt{fft} function is \texttt{void fft(int N, complex$^*$ X)}.

\stepcounter{figure}
\begin{lstlisting}[language=C,style=c,caption=Code to perform FFT on new frame,captionpos=b,firstline=371,lastline=377,label=fig:3.5.1]
    for (k=0;k<FFTLEN;k++)
	{                           
		inframe[k] = inbuffer[m] * inwin[k]; 
		if (++m >= CIRCBUF) m=0; /* wrap if required */
	} 
	
	/************************* DO PROCESSING OF FRAME  HERE **************************/
\end{lstlisting}

As this is a real-time implementation, optimizations are required to decrease the run time of frame processing. One of the primary optimizations is to only process half of the frame once in the frequency domain. As the frame being processed is real in the time domain, the frequency domain of the frame will be conjugate complex symmetric (\ref{eq:3.5.1}).

\begin{equation}
    X_{N-n} = X_{n}^*\label{eq:3.5.1}
\end{equation}

where $X_n$ is the value of the $N$ point FFT at frequency bin $n$ and $^*$ is the complex conjugate operator. 

Next,    the magnitude of the current frame is computed and used to implement the MMSE algorithm mentioned previously. 

\stepcounter{figure}
\begin{lstlisting}[language=C,style=c,caption=Computation of frame magnitude,captionpos=b,firstline=379,lastline=384,label=fig:3.5.2]
//N.B. most of the frame processing is done withing a for loop
//that takes advantage of the conjugate complex symmetry.
//This greatly improves efficiency.
for(k=0; k<FFTLEN/2; k++){
	//Calculate magnitude of current frame
	mag[k] = cabs(intermediate[k]);
\end{lstlisting}

\stepcounter{figure}
\begin{lstlisting}[language=C,style=c,caption=Implementation of MMSE algorithm (1),captionpos=b,firstline=405,lastline=406,label=fig:3.5.3]
//Check for possible MMSE elements in current frame
m1[k] = min(mag[k],m1[k]);
\end{lstlisting}

\stepcounter{figure}
\begin{lstlisting}[language=C,style=c,caption=Implementation of MMSE algorithm (2),captionpos=b,firstline=413,lastline=414,label=fig:3.5.4]
//Calculate MMSE from MMSE buckets
mmse[k] = min(min(m1[k],m2[k]),min(m3[k],m4[k]));
\end{lstlisting}

\stepcounter{figure}
\begin{lstlisting}[language=C,style=c,caption=Implementation of MMSE algorithm (3),captionpos=b,firstline=500,lastline=514,label=fig:3.5.5]
//Check if current MMSE bucket if full
if(++countMin >= BUCKET_FRAMES){
	countMin = 0; //Reset the counter
	
	//Reset oldest MMSE bucket
	for(k=0; k<FFTLEN/2; k++)
		m4[k] = mag[k];
	
	//Swap buckets
	temp = m4;
	m4 = m3;
	m3 = m2;
	m2 = m1;
	m1 = temp;
}
\end{lstlisting}

Finally, the value of $g(\omega)$ (\ref{eq:3.4.1}) for the current frequency bin is computed and elements of the \texttt{intermediate} array are overwritten accordingly. 

\stepcounter{figure}
\begin{lstlisting}[language=C,style=c,caption=Computation of $g(\omega)$ for single frequency bin ,captionpos=b,firstline=435,lastline=435,label=fig:3.5.6]
gw = max(lambda,(1-(alpha*mmse[k]/mag[k])));
\end{lstlisting}

\stepcounter{figure}
\begin{lstlisting}[language=C,style=c,caption=Overwriting the elements of \texttt{intermediate} ,captionpos=b,firstline=435,lastline=435,label=fig:3.5.7]
gw = max(lambda,(1-(alpha*mmse[k]/mag[k])));
\end{lstlisting}

It must be noted that once the IFFT of the \texttt{intermediate} array is taken, only the real values of the elements of \texttt{intermediate} will be written to \texttt{outframe} as any complex values will be due to finite precision effects.

\stepcounter{figure}
\begin{lstlisting}[language=C,style=c,caption=Writing the real values of \texttt{intermediate} to \texttt{outframe} ,captionpos=b,firstline=516,lastline=521,label=fig:3.5.8]
//Calculate the IFFT of the current frame
ifft(FFTLEN, intermediate);
	
//Copy real part of X[k] to the output frame
for(k=0; k<FFTLEN; k++)
	outframe[k] = intermediate[k].r;
\end{lstlisting}

\subsection{Performance of the Basic Implementation}
\begin{figure}[H]
\centering
\includegraphics[width=1\linewidth]{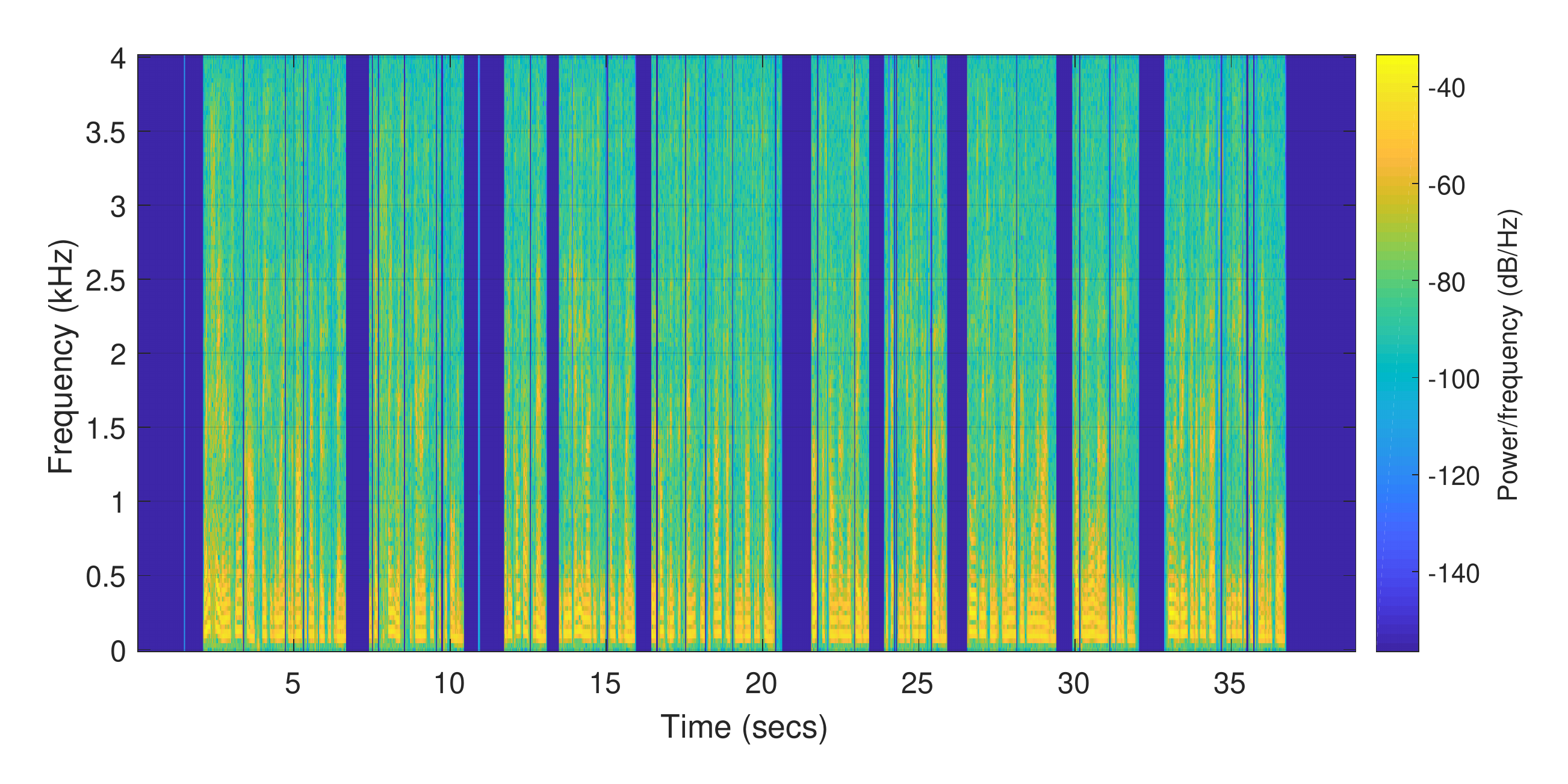}
\caption{Spectrogram of \texttt{clean.wav}}
\label{fig:3.6.1}
\end{figure}

\begin{figure}[H]
\centering
\includegraphics[width=1\linewidth]{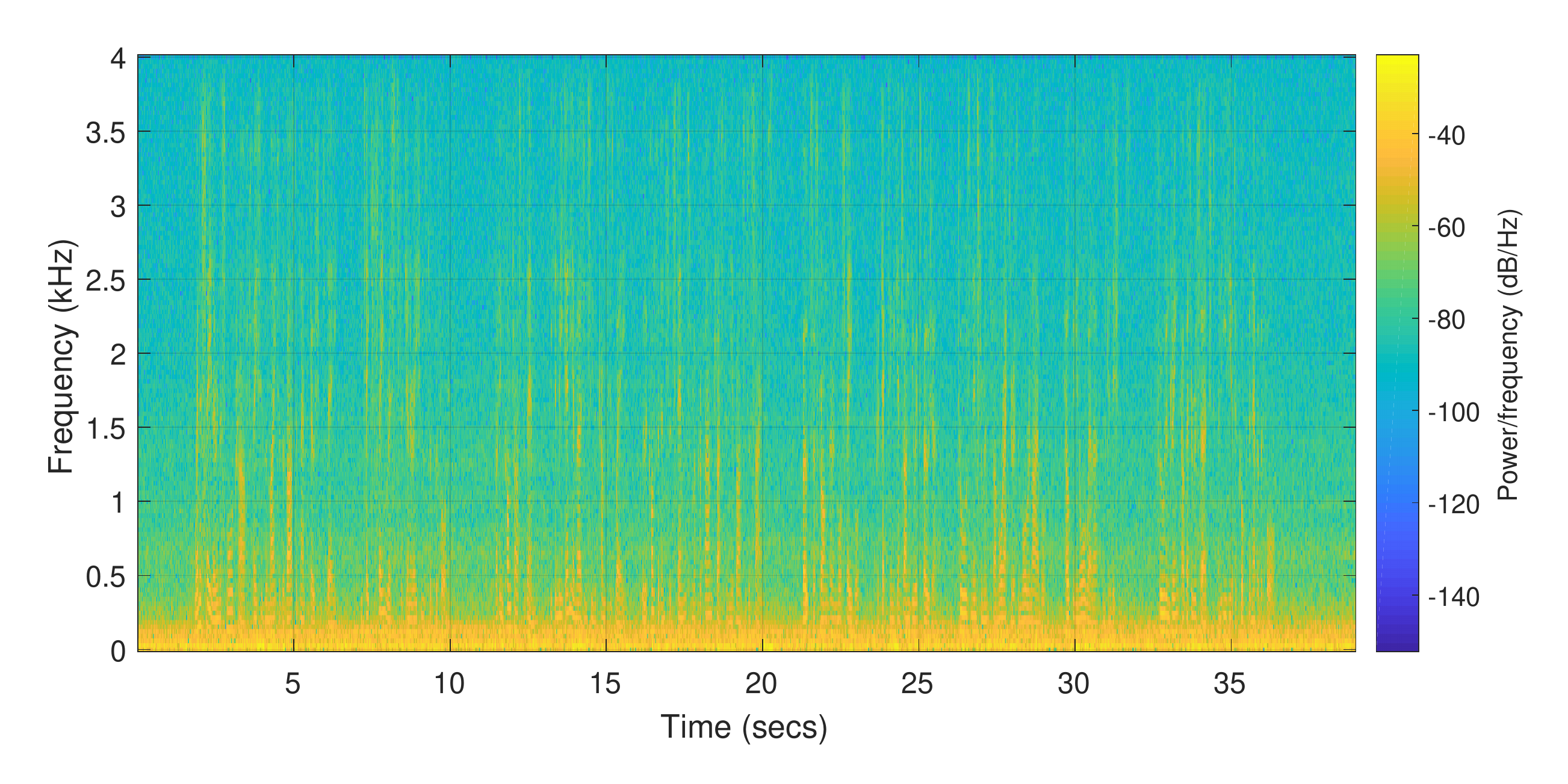}
\caption{Spectrogram of \texttt{car1.wav}}
\label{fig:3.6.2}
\end{figure}

\begin{figure}[H]
\centering
\includegraphics[width=1\linewidth]{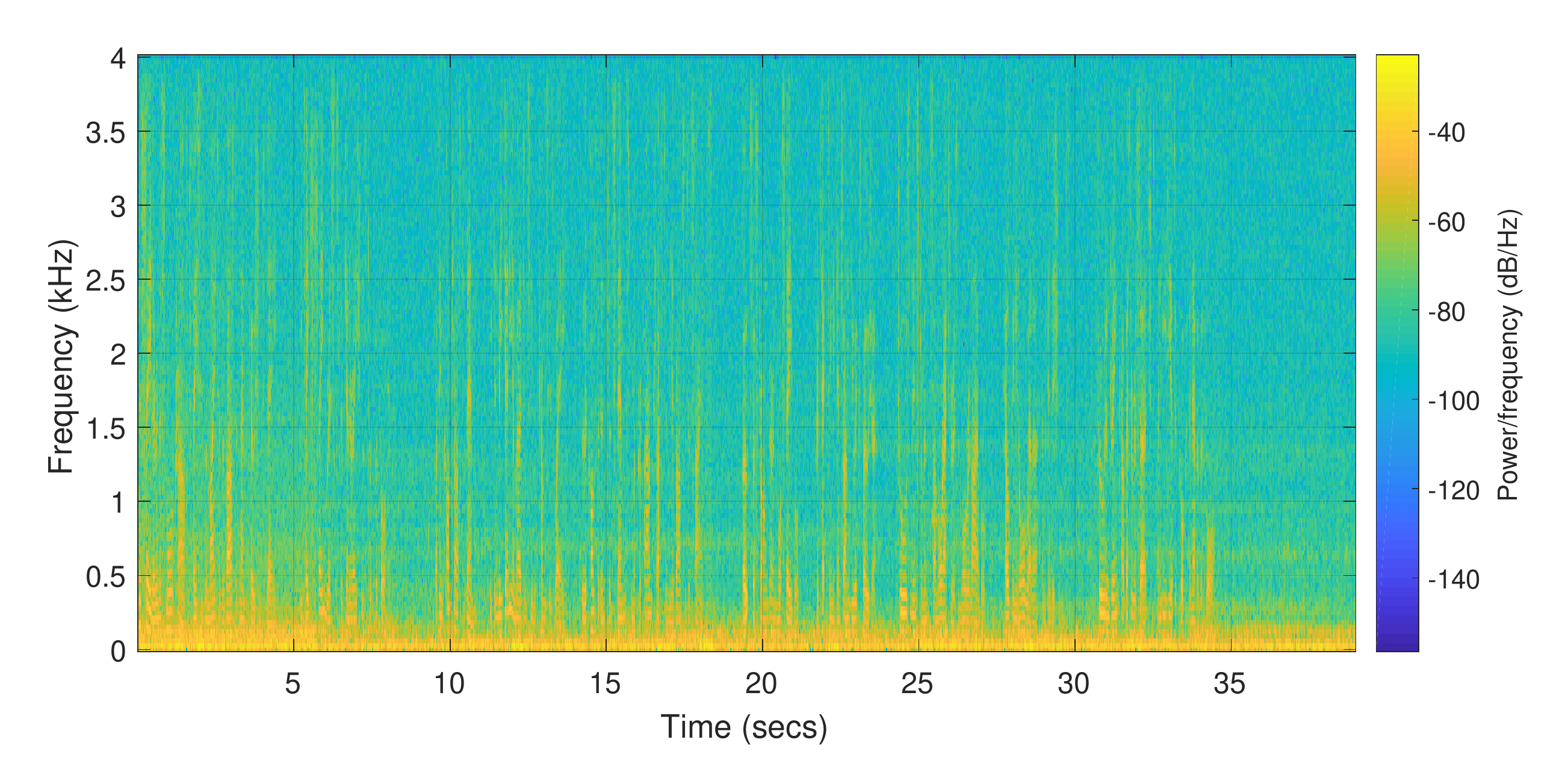}
\caption{Spectrogram of \texttt{car1.wav} output with $\alpha=20$ and $\lambda=0.05$ (Basic Implementation)}
\label{fig:3.6.3}
\end{figure}

The performance of this basic implementation will be used as a benchmark to compare the enhancements that will be introduced in the next section. To compare the different implementations, a selection of Waveform Audio Files (i.e. \texttt{.wav}) containing "the sailor passage" with different types of added noise (e.g. car, factory, helicopter) at different noise levels were used as input to the system. To refer to the different types of input their file names will be used (e.g. \texttt{phantom2.wav} for added noise from the F15 phantom aircraft at noise level 2). The spectrogram of the  input with no added noise (i.e. \texttt{clean.wav}) is shown in Figure \ref{fig:3.6.1}. The spectrogram of \texttt{car1.wav} in Figure \ref{fig:3.6.2}. Through a visual inspection of the spectrogram, the car noise seems to have added broadband stationary noise at low frequencies (i.e. less that 300Hz). The spectrogram of \texttt{car1.wav} after processing, which will simply be referred to as "the output," is shown in Figure \ref{fig:3.6.3}. As expected, the basic spectral subtraction implementation has reduced the noise in the signal; however, improvements can still be made.


\section{Enhancements}
In this sections various enhancements are made to the basic implementation. Not all enhancements were used in the final implementation as some proved to have little effect in practice given their computational cost. The C-code implementation for all of the enhancements can be found in the Appendix.

\subsection{Low-pass filtering the magnitude}
\begin{figure}[H]
\centering
\includegraphics[width=1\linewidth]{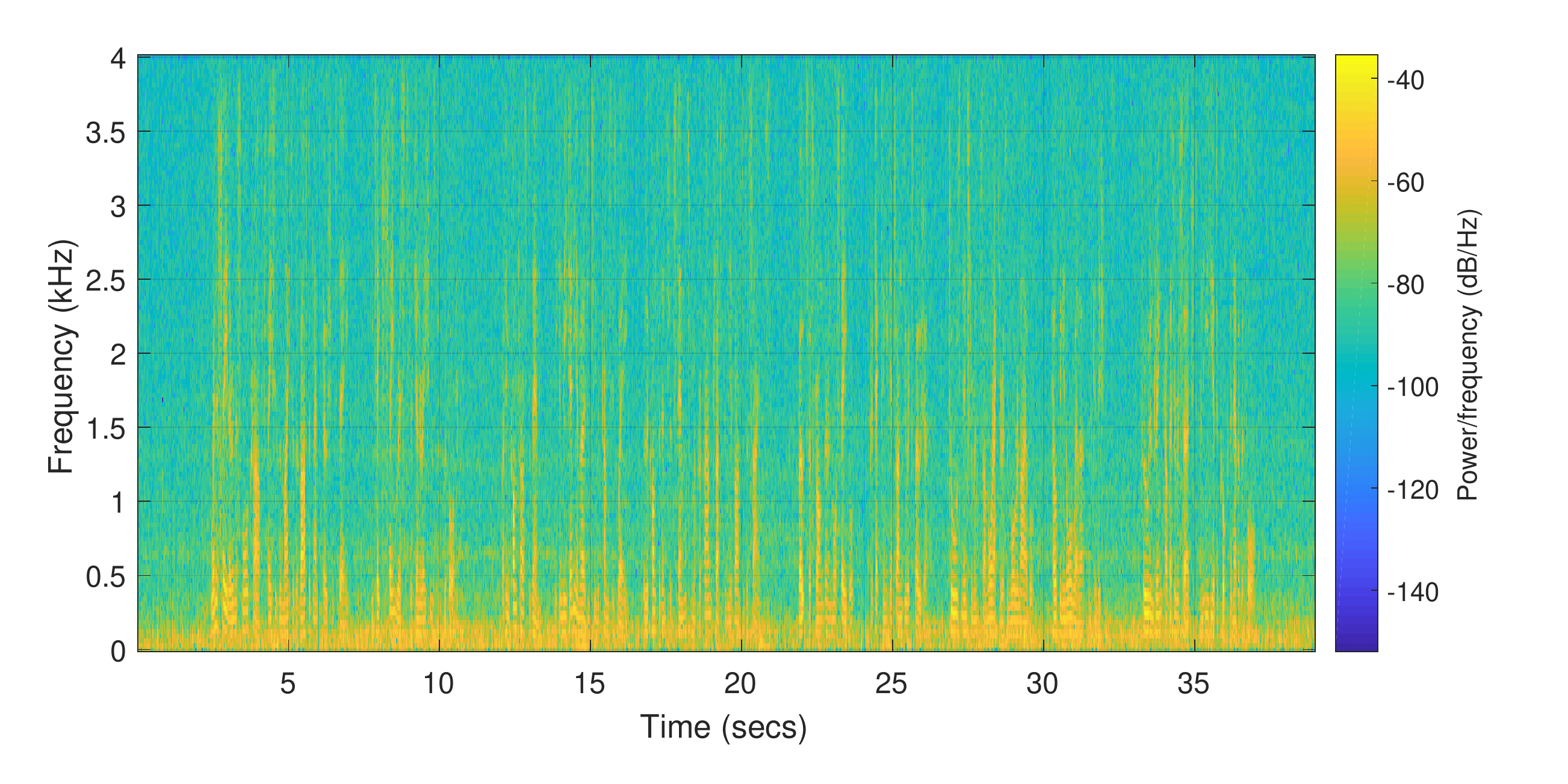}
\caption{Spectrogram \texttt{car1.wav} output with $\alpha=2$, $\lambda=0.05$, $\tau=0.03$ (Enhancement 1)}
\label{fig:4.1.1}
\end{figure}

The first enhancement is simply to low-pass filter the magnitude $\abs{X(\omega)}$ of the frame. Note the the low-pass filter in acting on consecutive frames rather than in the time domain. This was recommended in \cite{martin1994spectral} \cite{lockwood1992non}. The low-pass filtering is done according to the difference equation (\ref{eq:3.6.1})

\begin{equation}
    P_t(\omega) = (1 - k)\abs{X(\omega)} + kP_{t-1}(\omega) \label{eq:3.6.1}
\end{equation}

where $k = e^{T/\tau}$ is the z-plane pole for time constant $\tau$ and frame rate $T$ and $P_t(\omega)$ is the low-pass filtered input for frame $t$. Note the since $\abs{e^{T/\tau}} < 1$ for $T \neq 0$, the filter will always be stable for any value of $\tau$. This enhancement improved significantly the output while $\alpha$ was reduced from 20 to 2; $\tau$ was set empirically to 30ms which is in the range suggested by \cite{mitchProjNotes}. The spectrogram of the output with the above enhancement is shown in Figure \ref{fig:4.1.1}. Surprisingly, even though the spectrum looks similar to Figure \ref{fig:3.6.3}, it was perceived to be much clearer.

\subsection{Low-pass filtering power}
\begin{figure}[H]
\centering
\includegraphics[width=1\linewidth]{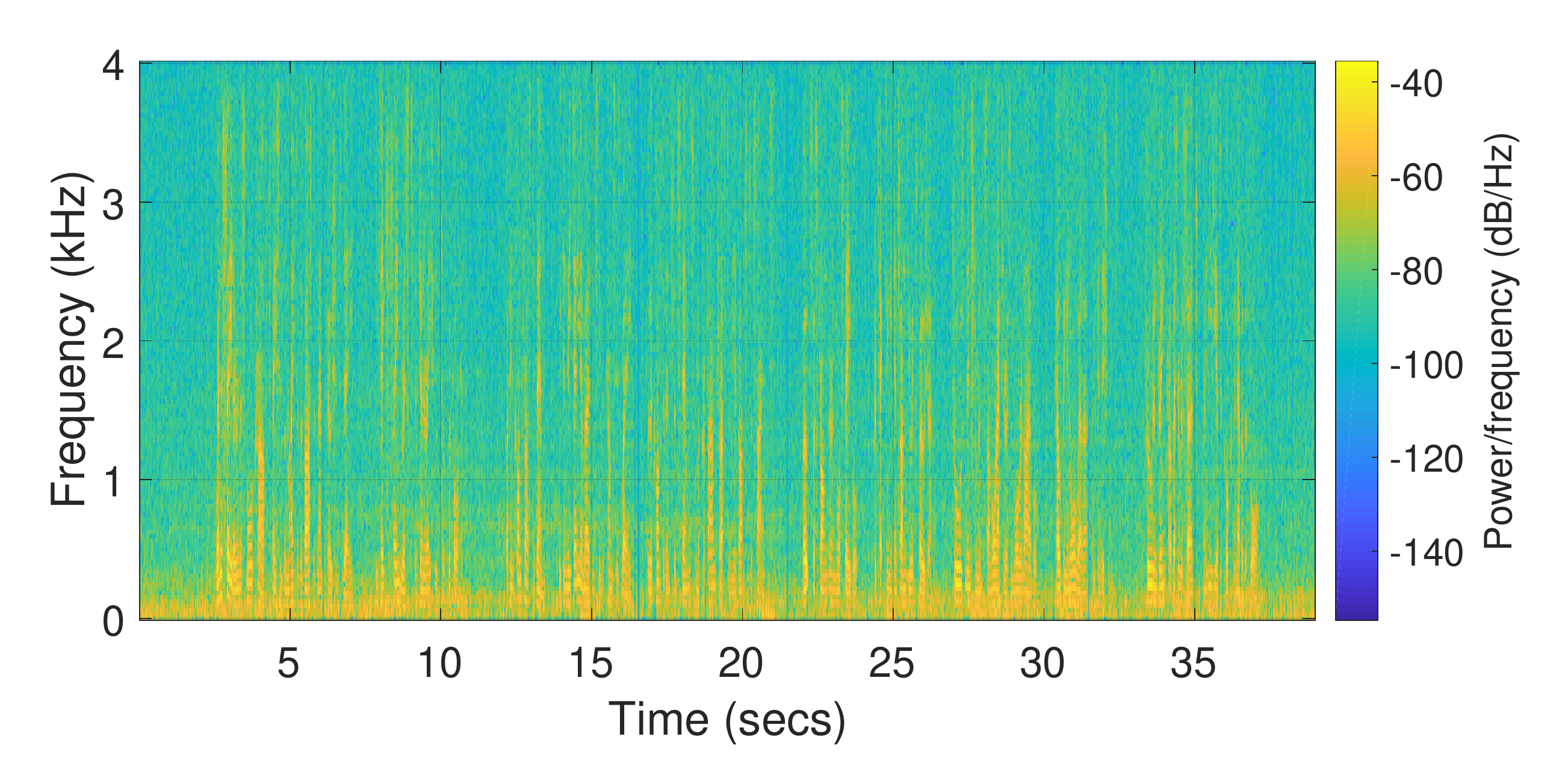}
\caption{Spectrogram \texttt{car1.wav} output with $\alpha=2$, $\lambda=0.05$, $\tau=0.025$ (Enhancement 2)}
\label{fig:4.2.1}
\end{figure}

This enhancement is very similar to enhancement 1, except instead of low-pass filtering the magnitude $\abs{X(\omega)}$, the power $\abs{X(\omega)}^2$ is low-pass filtered. Theoretically, this makes sense since humans perceive power rather than magnitude. Furthermore, it's expected that the optimal value of $\tau$ will decrease as $\abs{X(\omega)}^2$ will vary faster than $\abs{ X(\omega)}$. Empirically, the optimal value of $\tau$ was set to 0.025. This is within the range specified by \cite{mitchProjNotes}. The spectrogram of the output with the above enhancement is shown in Figure \ref{fig:4.2.1}. The output was perceived to be of higher quality than the output when using enhancement 1.

\subsection{Low-pass filtering the noise}
In this enhancement, instead of low-pass filtering the magnitude of the frame, the MMSE is low-pass filtered. Theoretically, robustness of the system to non-stationary nose where there would be a abrupt change in the output once the MMSE frames $M_i(\omega)$ shift. Empirically, there was a noticeable difference when the input to the DSK was set to \texttt{factory1.wav} and \texttt{factory2.wav} as they contain "the sailor" passage with added factory noises at different levels.

\subsection{Using different values for \texorpdfstring{$g(\omega)$}{Lg}}
This enhancement consists of implementing different versions of $g(\omega)$ shown below:

\begin{align}
    g(\omega) &= \text{max}\left(\lambda \frac{\abs{\hat{N}(\omega)}}{\abs{X(\omega)}}, 1-\alpha\frac{\abs{\hat{N}(\omega)}}{\abs{X(\omega)}}\right)\\
    g(\omega) &= \text{max}\left(\lambda \frac{\abs{P(\omega)}}{\abs{X(\omega)}}, 1-\alpha\frac{\abs{\hat{N}(\omega)}}{\abs{X(\omega)}}\right)\\
    g(\omega) &= \text{max}\left(\lambda \frac{\abs{\hat{N}(\omega)}}{\abs{P(\omega)}}, 1-\alpha\frac{\abs{\hat{N}(\omega)}}{\abs{P(\omega)}}\right)\label{eq:4.4.3}\\
    g(\omega) &= \text{max}\left(\lambda, 1-\alpha\frac{\abs{\hat{N}(\omega)}}{\abs{P(\omega)}}\right)
\end{align}

All of these enhancements were tested empirically, the best performing one was (\ref{eq:4.4.3}) which is also the version of $g(\omega)$ that is used in \cite{berouti1979enhancement}.

\subsection{Calculating \texorpdfstring{$g(\omega)$}{Lg} in the power domain}
This is yet another enhancement that modifies $g(\omega)$; however, in this case, the modification is different as $g(\omega)$ will be computed in the power domain instead of the magnitude domain ( \ref{eq:4.5.1}).

\begin{equation}
    g(\omega) = \text{max}\left(\lambda, \sqrt{1 - \left(a\frac{\abs{\hat{N}(\omega)}}{\abs{X(\omega)}}\right)^2}\right) \label{eq:4.5.1}
\end{equation}
As mentioned previously, humans perceive power rather than magnitude so there is theoretical justification for this enhancement. However, empirically, little difference was perceived in the output signal with this enhancement being very computationally expensive due to the \texttt{powf} and \texttt{sqrtf} functions that must be used.

\subsection{Overestimate \texorpdfstring{$\alpha$}{Lg} at lower SNR frequency bins}
\begin{figure}[H]
\centering
\includegraphics[width=1\linewidth]{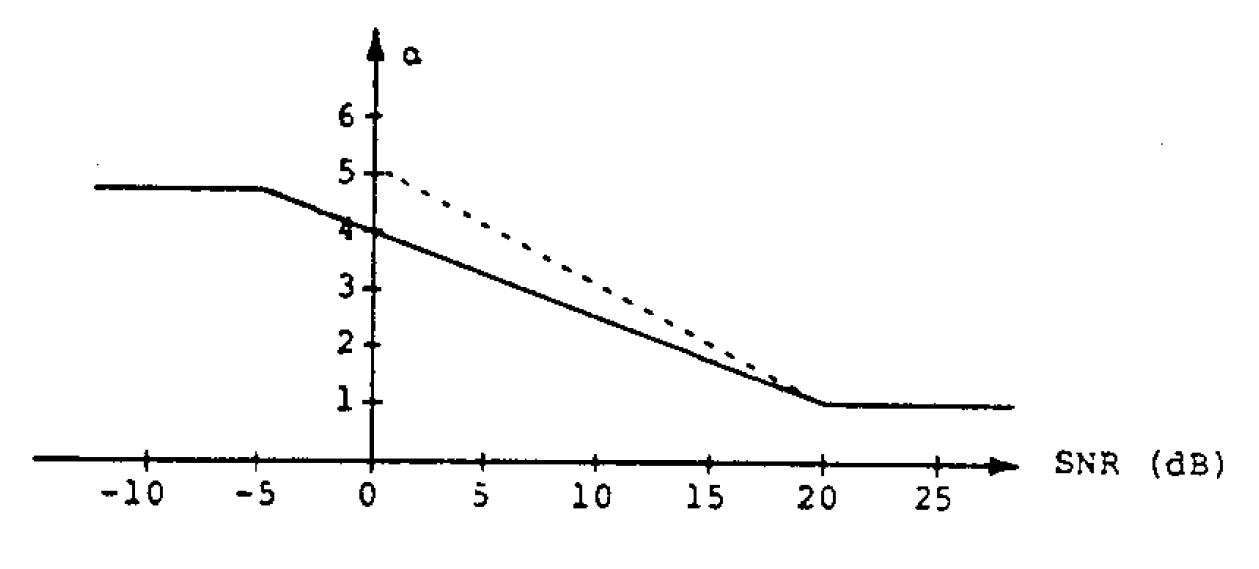}
\caption{Value of the compensation factor $\alpha$ versus SNR of frame}
\label{fig:4.6.1}
\end{figure}

This enhancements adjusts the parameter $\alpha$ from frame to frame depending on the SNR as suggested by \cite{berouti1979enhancement}. The SNR from frame to frame will vary as the power of the noise will be approximately the same for stationary noise while the power of the signal will vary. For high SNR frames, increasing the value of $\alpha$ is not necessary and will lead to a distortion in the speech signal. For low SNR frames, a higher value $\alpha$ is necessary to suppress the noise. Therefore, there is theoretical justification to this enhancement. As suggested by \cite{berouti1979enhancement}, a piece wise linear function was used to select the value of $\alpha$ (Figure \ref{fig:4.6.1}) (\ref{eq:4.6.1}).

The function for the solid line in Figure \ref{fig:4.6.1} is:

\begin{equation}
    \alpha(SNR) = \begin{cases} 5 & \text{for } SNR < -5 \\
                              5 - \frac{4}{20}SNR & \text{for } -5 \leq SNR \leq 20\\
                              1 & \text{for } SNR > 1
                  \end{cases} \label{eq:4.6.1}
\end{equation}

Even though in \cite{berouti1979enhancement}, this enhancement was only performed with a frame by frame granularity (i.e. the value of $\alpha$ will change from frame to frame; however, it will remain constant within a frame) the enhancement was further modified to allow for $\alpha$ to change with a frequency bin granularity which was used by \cite{kamath2002multi}. The justification of this is that noise does not effect the speech signal in the frequency domain uniformly. This is illustrated in Figure \ref{fig:4.6.1} which shows the SNR ratios of four linearly space frequency bins across consecutive frames for the input corrupted by the added car noise. Bin 1 has a lower SNR across most frames as it corresponds to the very low frequencies ( $< 10$Hz) which is where the car noise is mostly present. The SNRs between different frequency bins differ substantially with difference being greater than 100dB for some frames. Note the if the slope of the piece-wise linear function (\ref{eq:4.6.1}) is increased, then the temporal dynamic range of the signal will also increase substantially leading to a distorted output. Empirically, the slope used in \cite{berouti1979enhancement}, was confirmed to have a good performance so it was not modified.

\begin{figure}[H]
\centering
\includegraphics[width=1\linewidth]{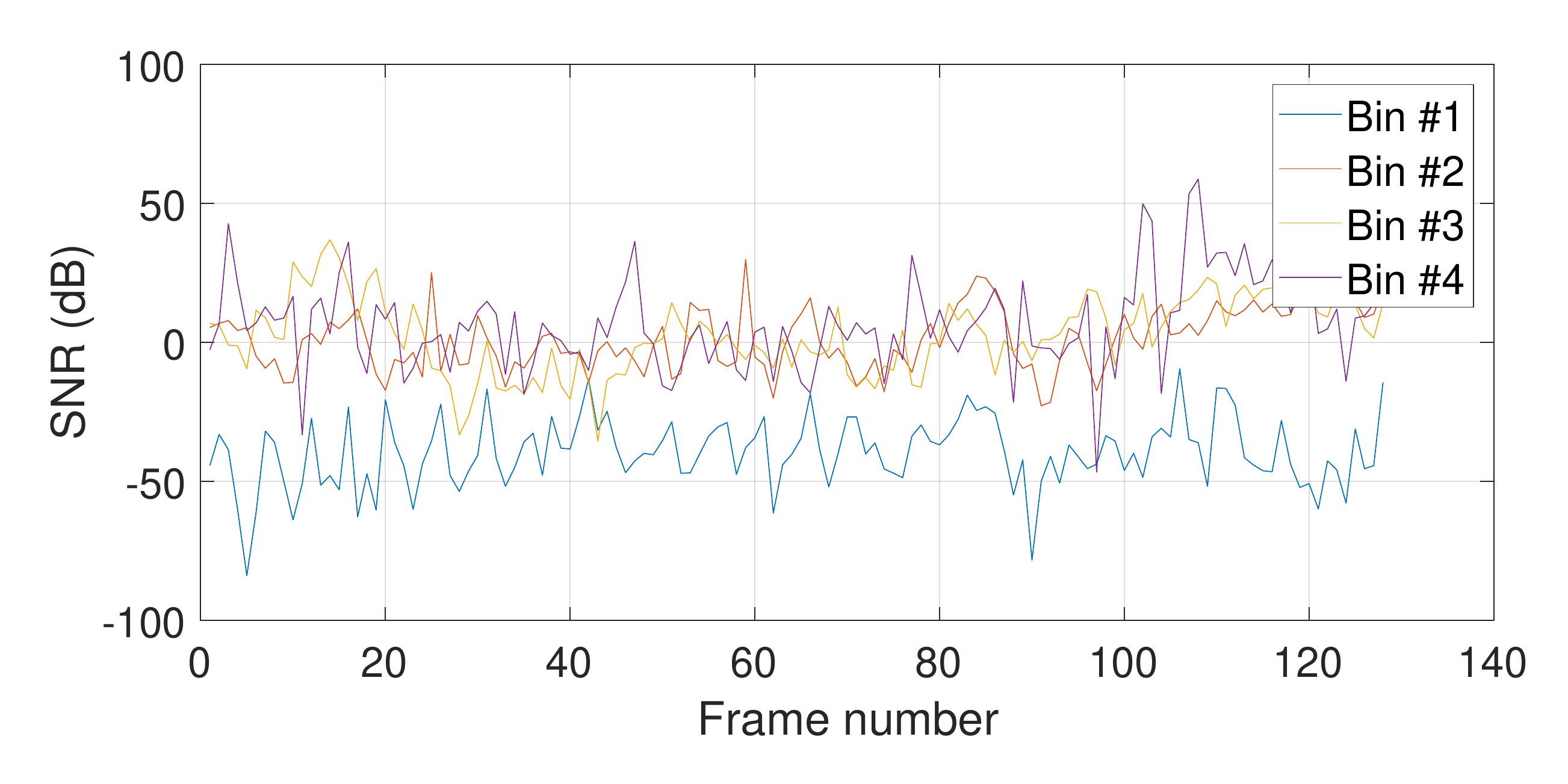}
\caption{SNR ratios of four linearly spaced frequency bins across consecutive frames}
\label{fig:4.6.2}
\end{figure}

\subsection{Adding the \texorpdfstring{$\delta(F)$}{Lg} term}
\begin{figure}[H]
\centering
\includegraphics[width=1\linewidth]{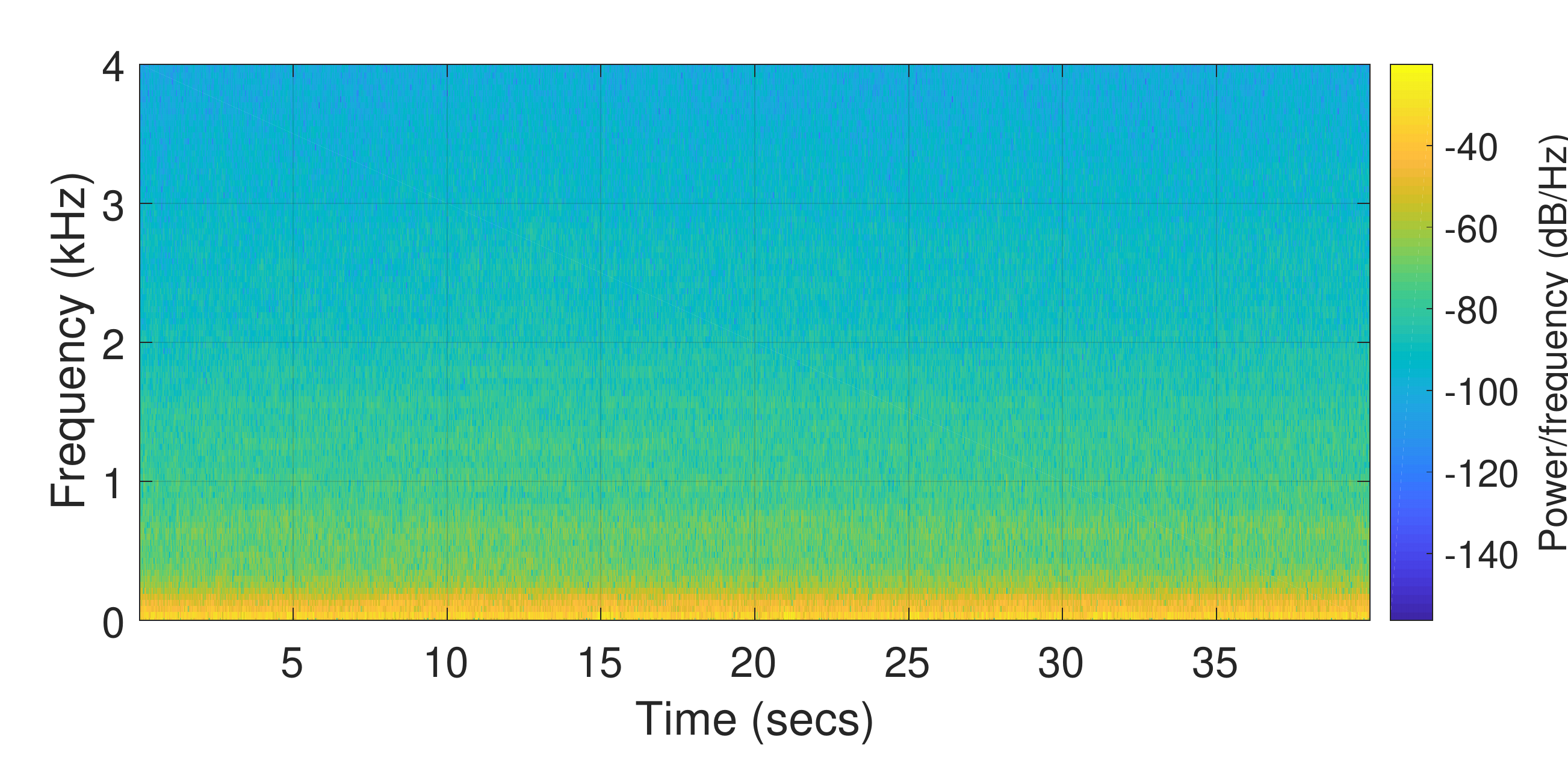}
\caption{Spectrogram of added car noise in \texttt{car1.wav}}
\label{fig:4.7.1}
\end{figure}

In addition to adjusting the noise estimate, $\hat{N}(\omega)$ based on the SNR ratio of each frequency bin, this enhancement aims to further adjust  $\hat{N}(\omega)$ based the analogue frequency, $F$ that the frequency bins represents. This enhancement was introduced by \cite{kamath2002multi} and uses a "tweaking factor" $\delta(F)$ that can be individually set for each frequency bin. In the real-world, noise (e.g. added car noise) is coloured and affects certain frequencies more than others. This is illustrated in Figure \ref{fig:4.7.1} which shows the spectrogram of the added car noise. Note that the car noise is present primarily at frequencies $0\text{Hz} < F < 300$Hz which explains the discrepancies between the SNRs of different frequency bins in Figure {\ref{fig:4.6.2}}.

The $\delta(F)$ terms adds an additional degree of freedom to the noise subtraction level of each frequency and modifies (\ref{eq:3.4.1}) slightly to the form shown in (\ref{eq:4.7.1})

\begin{equation}
    g(\omega) = \text{max}\left(\lambda,1 - \delta(F)\alpha(SNR)\frac{\abs{\hat{N}(\omega)}}{\abs{X(\omega)}}\right)\label{eq:4.7.1}
\end{equation}

The values of $\delta(F)$ where determined empirically and set to:

\begin{equation}
    \delta(F) = \begin{cases} 1     & 0Hz < F < 1kHz\\
                              2.5   & 1kHz \leq F < 2kHz\\
                              1.5   & 2kHz \leq F 
                \end{cases}\label{eq:4.7.2}
\end{equation}

\begin{figure}[H]
\centering
\includegraphics[width=1\linewidth]{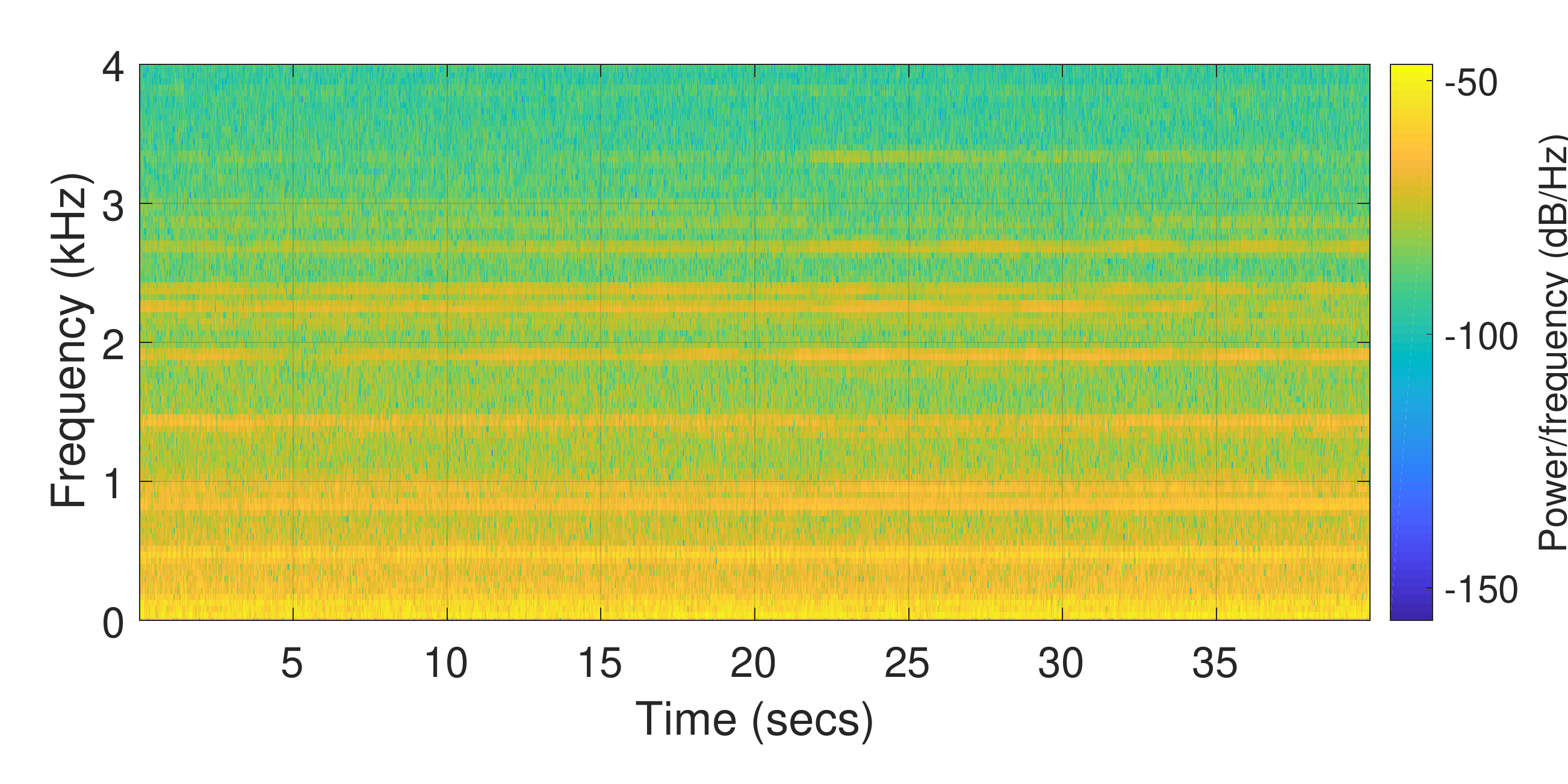}
\caption{Spectrogram of added helicopter noise in \texttt{lynx1.wav}}
\label{fig:4.7.2}
\end{figure}

These values match the ones used in \cite{kamath2002multi}. The addition of the delta term, lead to a significant increase in the intelligibility of the output especially when dealing with added helicopter noise. The spectrogram of added helicopter noise is shown in Figure \ref{fig:4.7.2}

\begin{figure}[H]
\centering
\includegraphics[width=1\linewidth]{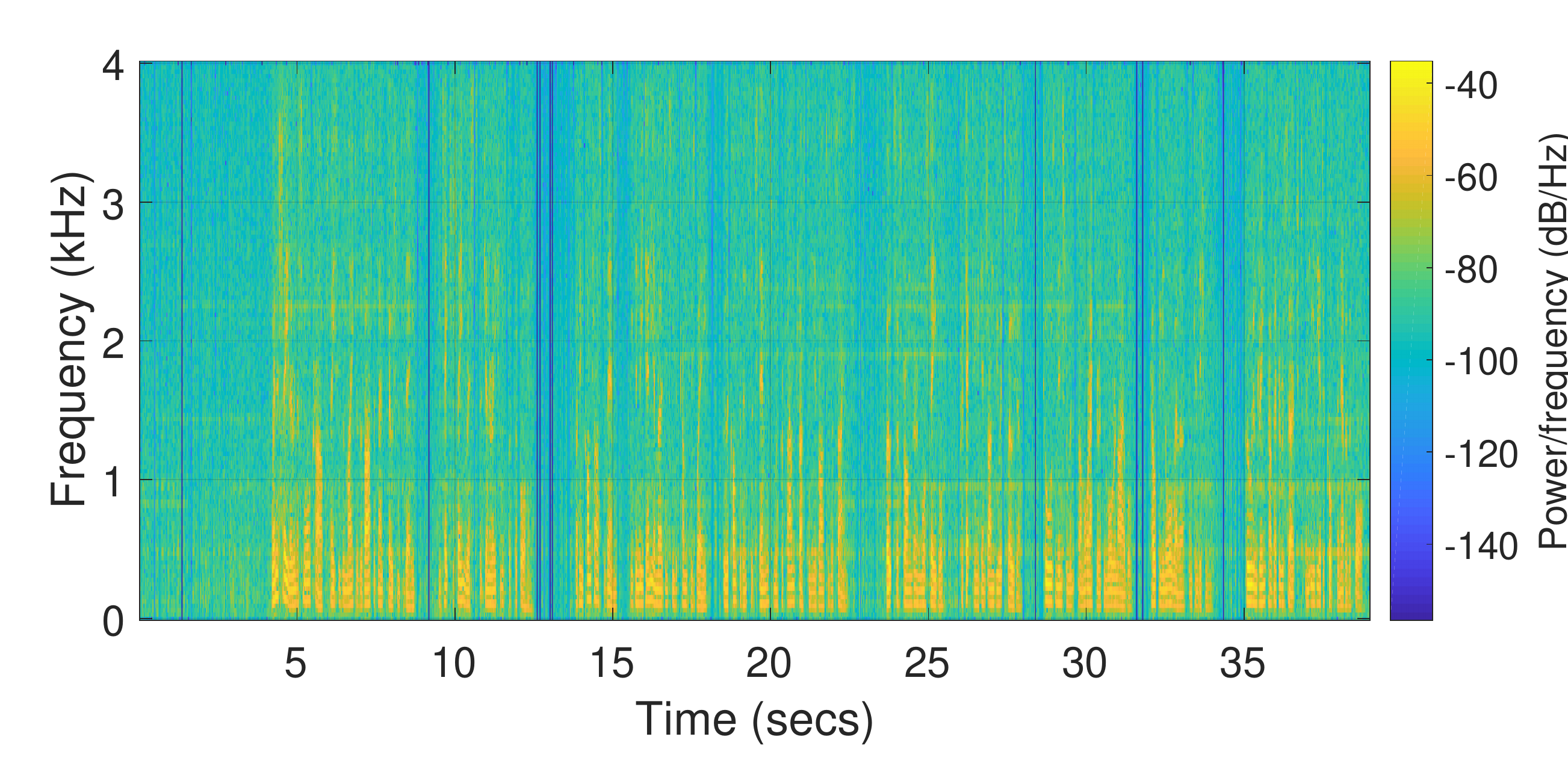}
\caption{Spectrogram of \texttt{lynx1.wav} output with $\delta(F)$ set to (\ref{eq:4.7.2}) (Enhancement 7)}
\label{fig:4.7.3}
\end{figure}

\subsection{Using different frame lengths}
By changing the frame length, the time and frequency resolution of the implementation can be changed. A larger frame length will effectively increase the frequency resolution of each frame while decreasing the frequency resolution and vice versa. As mentioned previously, the basic implementation had a frame length of 256 samples and a sampling frequency of 8kHz. Thus each frame consists of 32ms of speech. A shorter frame length resulted in "roughness" in the speech while a longer frame length lead to "slurred" speech. These results agree with \cite{mitchProjNotes}. Overall, the ideal frame length was found to be around 28ms. The frame length was adjusted by changing the \texttt{FFTLEN} definition in the C code. 

\subsection{Residual Noise Reduction}
This enhancement attempts to remove some of the musical noise by taking advantage of the frame to frame randomness \cite{boll1979suppression}. Effectively, as mentioned previously, the musical noise is due to the formation of peaks in the magnitude spectrum which will appear at a random amplitude and frequency for each frame. Therefore, the musical noise can be suppressed by replacing the frequency bins of the current frame with the minimum frequency bins from the previous and next frame.

\begin{equation}
    \abs{X_i(\omega)} = \text{min}\left(\abs{X_{i-1}(\omega)},\abs{X_i(\omega)},\abs{X_{i+1}(\omega)}\right)\label{eq:4.8.1}
\end{equation}

However, even with the complex conjugate symmetric optimization, this enhancement is very computationally demanding and could not be implemented in parallel with the enhancements mentioned thus far. For this reason, it was not included in the final implementation.

\subsection{Reduce the MMSE Memory}
This enhancement aims to increase the responsiveness of the system to non-stationary noise by reducing the MMSE memory. Reducing the MMSE memory is also beneficial from a computational point of view; however, if the speaker continues to produce sound for more than the MMSE memory (measured in seconds), the noise estimate that will be made will be extremely high as segments of speech have effectively been misclassified as noise. This will lead to a serious distortion in the speech signal.

\subsection{Changing the windowing function}
A final enhancement that was considered was to use a different windowing function. As mentioned in section 3.2, in the implementations thus far, the Hamming window was used to mitigate the effects of spectral artifacts in the frequency domain. Other windows that were considered were the Hanning, Gaussian and Black-Harris (3-term). Out of these windows, the Hanning performed the best which might be due to it's higher spectral roll-off (Figure \ref{fig:3.2.3})

\section{Final Implementation and Results}
\begin{figure}[H]
\centering
\includegraphics[width=1.1\linewidth]{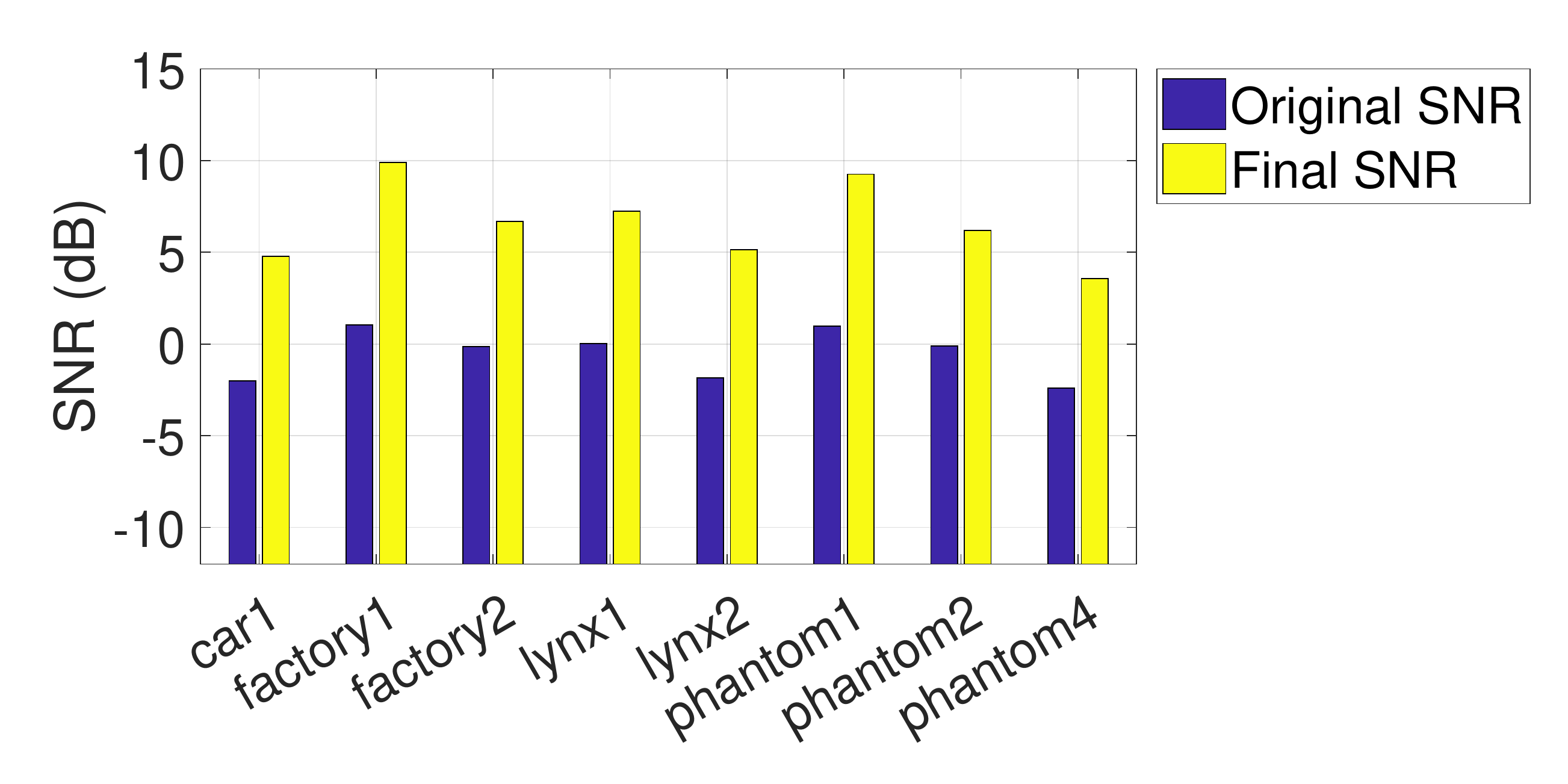}
\caption{Improvement in SNR levels with final implementation}
\label{fig:5.1.1}
\end{figure}

In the final implementation a compromise between computational complexity and system performance was made when choosing which enhancements to include. Enhancements 4.2, 4.3, 4.4, 4.6, 4.7 and 4.11 were included in the final implementation. Enhancement 4.5 and 4.9 were very computationally demanding and could not be included together with other enhancements while the rest of the enhancements did not improve the final output or, in some case, lead to worse performance. The input and output SNR levels for the final implementation is shown in Figure \ref{fig:5.1.1}. The final implementation managed to reduce the noise significantly for all inputs; however, it performs best when the original signal has a high original SNR level. It had the worse improvement in SNR with the \texttt{phantom4.wav} input were it only managed to achieved a 5.98dB improvement.

\section{Conclusion}
A real-time speech enhancement system was implemented based on the spectral subtraction technique. Different enhancements were considered and their performance was evaluated based on extensive listening tests, spectrograms and SNR comparisons. The final system manages to reduce the noise present in the output signal substantially while achieving a compromise between broadband noise, musical noise and speech intelligibility. Nevertheless, the system struggles to deal with very low SNR inputs. To deal with these types of inputs, other more recent noise reduction techniques such as Wiener filters or signal subspace approaches could be used.

%Add References to table of contents
\addcontentsline{toc}{section}{References}
\printbibliography

@misc{princOfDigCom,
    title={6.450 Principles of Digital Communications I.},
    chapter={Lecture Notes: Random processes and noise},
    author={Lizhong Zhen and Robert Gallager},
    year={2006},
    note = {Accessed: 2019-03-07},
    organization={Massachusetts Institute of Technology: MIT OpenCourseWare},
    pages={203-204},
    URL={https://ocw.mit.edu/courses/electrical-engineering-and-computer-science/6-450-principles-of-digital-communications-i-fall-2006/lecture-notes/book_7.pdf}
}

@misc{mitchProjNotes,
    title={Project: Speech Enhancement},
    author={Paul D. Mitcheson},
    organization={Imperial College London},
    edition={Version 1.3}
}

@article{boll1979suppression,
  title={Suppression of acoustic noise in speech using spectral subtraction},
  author={Boll, Steven},
  journal={IEEE Transactions on acoustics, speech, and signal processing},
  volume={27},
  number={2},
  pages={113--120},
  year={1979},
  publisher={IEEE}
}

@article{martin1994spectral,
  title={Spectral subtraction based on minimum statistics},
  author={Martin, Rainer},
  journal={power},
  volume={6},
  pages={8},
  year={1994}
}

@inproceedings{berouti1979enhancement,
  title={Enhancement of speech corrupted by acoustic noise},
  author={Berouti, Michael and Schwartz, Richard and Makhoul, John},
  booktitle={ICASSP'79. IEEE International Conference on Acoustics, Speech, and Signal Processing},
  volume={4},
  pages={208--211},
  year={1979},
  organization={IEEE}
}

@inproceedings{lockwood1992non,
  title={Non-linear spectral subtraction (NSS) and hidden Markov models for robust speech recognition in car noise environments},
  author={Lockwood, P and Boudy, J and Blanchet, M},
  booktitle={[Proceedings] ICASSP-92: 1992 IEEE International Conference on Acoustics, Speech, and Signal Processing},
  volume={1},
  pages={265--268},
  year={1992},
  organization={IEEE}
}

@book{hartmann2004signals,
  title={Signals, sound, and sensation},
  author={Hartmann, William M},
  year={2004},
  publisher={Springer Science \& Business Media},
  page={256}
}

@inproceedings{kamath2002multi,
  title={A multi-band spectral subtraction method for enhancing speech corrupted by colored noise.},
  author={Kamath, Sunil and Loizou, Philipos},
  booktitle={ICASSP},
  volume={4},
  pages={44164--44164},
  year={2002},
  organization={Citeseer}
}

@article{spatio,
author = {Ioannides, Georgios and Kourouklides, Ioannis and Astolfi, Alessandro},
year = {2022},
month = {02},
pages = {2896},
title = {Spatiotemporal dynamics in spiking recurrent neural networks using modified-full-FORCE on EEG signals},
volume = {12},
journal = {Scientific Reports},
doi = {10.1038/s41598-022-06573-1}
}
\end{multicols}
%\newpage

%Add Appendix to table of contents
\addcontentsline{toc}{section}{Appendix}
\section*{Appendix}
\setcounter{figure}{1}
%Replace listings counter with figure counter and make it A,B,C
\renewcommand{\thelstlisting}{\Alph{figure}}
\begin{Huge}
\begin{lstlisting}[language=C,style=c,caption=C source code,captionpos=b,label=fig:a]
/*************************************************************************************
			       DEPARTMENT OF ELECTRICAL AND ELECTRONIC ENGINEERING
					   		     IMPERIAL COLLEGE LONDON 

 				      EE 3.19: Real Time Digital Signal Processing
					       Dr Paul Mitcheson and Daniel Harvey

				        		 PROJECT: Frame Processing

 				            ********* ENHANCE. C **********
							 Shell for speech enhancement 

  		Demonstrates overlap-add frame processing (interrupt driven) on the DSK. 

 *************************************************************************************
 				             By Danny Harvey: 21 July 2006
							 Updated for use on CCS v4 Sept 2010
 ************************************************************************************/
/*
 *	You should modify the code so that a speech enhancement project is built 
 *  on top of this template.
 */
/**************************** Pre-processor statements ******************************/
//  library required when using calloc
#include <stdlib.h>
//  Included so program can make use of DSP/BIOS configuration tool.  
#include "dsp_bios_cfg.h"

/* The file dsk6713.h must be included in every program that uses the BSL.  This 
   example also includes dsk6713_aic23.h because it uses the 
   AIC23 codec module (audio interface). */
#include "dsk6713.h"
#include "dsk6713_aic23.h"

// math library (trig functions)
#include <math.h>

/* Some functions to help with Complex algebra and FFT. */
#include "cmplx.h"      
#include "fft_functions.h"  

// Some functions to help with writing/reading the audio ports when using interrupts.
#include <helper_functions_ISR.h>

#define WINCONST 0.85185			/* 0.46/0.54 for Hamming window */
#define FSAMP 8000.0		/* sample frequency, ensure this matches Config for AIC */
#define FFTLEN 256					/* fft length = frame length 256/8000 = 32 ms*/
#define NFREQ (1+FFTLEN/2)			/* number of frequency bins from a real FFT */
#define OVERSAMP 4					/* oversampling ratio (2 or 4) */  
#define FRAMEINC (FFTLEN/OVERSAMP)	/* Frame increment */
#define CIRCBUF (FFTLEN+FRAMEINC)	/* length of I/O buffers */

#define OUTGAIN 16000.0				/* Output gain for DAC */
#define INGAIN  (1.0/16000.0)		/* Input gain for ADC  */

// PI defined here for use in your code 
#define PI 3.141592653589793
#define TFRAME (FRAMEINC/FSAMP)       /* time between calculation of each frame */

//Number of frames in a Minimum Magnitude Spectrum Estimate Bucket (MMSE Bucket)
#define BUCKET_FRAMES 312

//Define constants to select noise removal algorithm
//LOW_PASS_MAGNITUDE is used to Enable (1) or Disable (0) the low-pass filtering
//of the spectral magnitude. Please see report for more information.
#define LOW_PASS_MAGNITUDE 1
//LOW_PASS_NOISE is used to Enable(1) or Disable(0) the low-pass filtering
//of the Minimum Magnitude Spectral Estimate (MMSE). The MMSE is an estimator for the noise
//in the signal. Please see report for more information.
#define LOW_PASS_NOISE 0
//USE_MAGNITUDE_SQUARED is used to Enable(1) or Disable(0) the computation of
//square magnitude (i.e. spectral power) before applying the low-pass filer.
#define USE_MAGNITUDE_SQUARED 1
//Change the value of alpha at runtime according to the SNR of the signal
#define USE_DYNAMIC_ALPHA 1
//Select different G(w) to use.For values of G_OMEGA >= 3 some 
//prerequisits apply. If these have not been selected by the user, 
//the compilation will fail and an approriate error is 
//presented. This is used to improve robustness.
#define G_OMEGA 3
//USE_RESIDUAL_NOISE_REDUCTION is used to Enable(1) or Disable(0) the
//residual noise reduction algorithm as a way to remove the musical noise.
#define RESIDUAL_NOISE_REDUCTION 0
//DELTA_TERM is used to Enable(1) or Disable(0) the additional delta term
//for adjusting the noise estimator based on the frequency bin
#define DELTA_TERM 1

/******************************* Global declarations ********************************/

/* Audio port configuration settings: these values set registers in the AIC23 audio 
   interface to configure it. See TI doc SLWS106D 3-3 to 3-10 for more info. */
DSK6713_AIC23_Config Config = { \
			 /**********************************************************************/
			 /*   REGISTER	            FUNCTION			      SETTINGS         */ 
			 /**********************************************************************/\
    0x0017,  /* 0 LEFTINVOL  Left line input channel volume  0dB                   */\
    0x0017,  /* 1 RIGHTINVOL Right line input channel volume 0dB                   */\
    0x01f9,  /* 2 LEFTHPVOL  Left channel headphone volume   0dB                   */\
    0x01f9,  /* 3 RIGHTHPVOL Right channel headphone volume  0dB                   */\
    0x0011,  /* 4 ANAPATH    Analog audio path control       DAC on, Mic boost 20dB*/\
    0x0000,  /* 5 DIGPATH    Digital audio path control      All Filters off       */\
    0x0000,  /* 6 DPOWERDOWN Power down control              All Hardware on       */\
    0x0043,  /* 7 DIGIF      Digital audio interface format  16 bit                */\
    0x008d,  /* 8 SAMPLERATE Sample rate control        8 KHZ-ensure matches FSAMP */\
    0x0001   /* 9 DIGACT     Digital interface activation    On                    */\
			 /**********************************************************************/
};

// Codec handle:- a variable used to identify audio interface  
DSK6713_AIC23_CodecHandle H_Codec;

float *inbuffer, *outbuffer;   		/* Input/output circular buffers */
float *inframe, *outframe;          /* Input and output frames */
float *inwin, *outwin;              /* Input and output windows */
float ingain, outgain;				/* ADC and DAC gains */ 
float cpufrac; 						/* Fraction of CPU time used */
volatile int io_ptr=0;              /* Input/ouput pointer for circular buffers */
volatile int frame_ptr=0;           /* Frame pointer */

//Inermediate frame on which processing is done
complex *intermediate;

//Minimum Magnitude Spectrum Estimate (MMSE) Buckets
float* m1;
float* m2;
float* m3;
float* m4;

//Magnitude
float *mag;

//Parameters of g(w)
float alpha = 20;//16;
float lambda = 0.05;

//MMSE
float* mmse; 

#if LOW_PASS_MAGNITUDE == 1
	//Low-pass filtered version of mag(X) or mag(X)^2
	//N.B. This array is used to store both the previous frames
	//low-pass filtered estimate and the current low-pass filtered
	//estimate. This saves memory
	float* lpfMag;

	//Parameter used for low-pass filtering mag(X)
	float tau = 30e-3;
#endif 

#if USE_MAGNITUDE_SQUARED == 1
	//Low-pass filtered estimate of mag(X)^2
	float* lpfMagSquared;
#endif

#if LOW_PASS_NOISE == 1
	//Parameter used for low-pass filtering mmse (i.e. mag(N))
	float tauNoise = 80e-3;
#endif

#if USE_DYNAMIC_ALPHA == 1
	//Parameters for scalling of alpha with SNR
	float alphaMax = 10;
	float alphaMin = 1;
	float a0 = 6;
	float s = 0.001;
#endif

#if RESIDUAL_NOISE_REDUCTION == 1
	//Buffers for previous, current and next magnitude spectrum
	//Y_Next does not have to be allocated as Y_Next will be the value
	//currently calculated
	complex* Y_Prev;
	complex* Y_Curr;
#endif


 /******************************* Function prototypes *******************************/
void init_hardware(void);    	/* Initialize codec */ 
void init_HWI(void);            /* Initialize hardware interrupts */
void ISR_AIC(void);             /* Interrupt service routine for codec */
void process_frame(void);       /* Frame processing routine */

//Returns max of two floats 
float max(const float a, const float b);
//Returns min of two floats
float min(const float a, const float b);

/********************************** Main routine ************************************/
void main()
{      
  	int k; // used in various for loops
  
/*  Initialize and zero fill arrays */  

	inbuffer	= (float *) calloc(CIRCBUF, sizeof(float));	/* Input array */
    outbuffer	= (float *) calloc(CIRCBUF, sizeof(float));	/* Output array */
	inframe		= (float *) calloc(FFTLEN, sizeof(float));	/* Array for processing*/
    outframe	= (float *) calloc(FFTLEN, sizeof(float));	/* Array for processing*/
    inwin		= (float *) calloc(FFTLEN, sizeof(float));	/* Input window */
    outwin		= (float *) calloc(FFTLEN, sizeof(float));	/* Output window */
    
    //Frequency domain
    intermediate = (complex *) calloc(FFTLEN, sizeof(complex));
    
    //MMSE Buckets
    m1 = (float *) calloc(FFTLEN/2, sizeof(float));
    m2 = (float *) calloc(FFTLEN/2, sizeof(float));
    m3 = (float *) calloc(FFTLEN/2, sizeof(float));
    m4 = (float *) calloc(FFTLEN/2, sizeof(float));
    
    //Initialize memory to FLT_MAX (i.e. max value that float can take)
    for(k=0; k<FFTLEN/2; ++k){
    	m1[k] = FLT_MAX;
    	m2[k] = FLT_MAX;
    	m3[k] = FLT_MAX;
    	m4[k] = FLT_MAX;
    }
    
    //Magnitude of frequency domain
    mag = (float *) calloc(FFTLEN/2, sizeof(float));
    
    #if LOW_PASS_MAGNITUDE == 1
	    //Low-pass filtered estimate of current/previous frame
		lpfMag = (float *) calloc(FFTLEN/2, sizeof(float));
	#endif
    
    #if USE_MAGNITUDE_SQUARED == 1
		//Declare point to array to hold the squared magnitude values
		lpfMagSquared = (float *) calloc(FFTLEN/2, sizeof(float));
	#endif
	
	#if RESIDUAL_NOISE_REDUCTION == 1
	    //Define arrays for previous, current and next magnitude spectrum.
	    //Y_Next does not have to be allocated as Y_Next will be the value
		//currently calculated
	    Y_Prev = (complex *) calloc(FFTLEN, sizeof(complex));
	    Y_Curr = (complex *) calloc(FFTLEN, sizeof(complex));
	#endif
	
    
    //Allocating memory (i.e. defining) memory for 
	/* initialize board and the audio port */
  	init_hardware();
  
  	/* initialize hardware interrupts */
  	init_HWI();    
  
/* initialize algorithm constants */  
                       
  	for (k=0;k<FFTLEN;k++)
	{                           
	inwin[k] = sqrt((1.0-WINCONST*cos(PI*(2*k+1)/FFTLEN))/OVERSAMP);
	outwin[k] = inwin[k]; 
	} 
  	ingain=INGAIN;
  	outgain=OUTGAIN;
  							
  	/* main loop, wait for interrupt */  
  	while(1) 	process_frame();
}
    
/********************************** init_hardware() *********************************/  
void init_hardware()
{
    // Initialize the board support library, must be called first 
    DSK6713_init();
    
    // Start the AIC23 codec using the settings defined above in config 
    H_Codec = DSK6713_AIC23_openCodec(0, &Config);

	/* Function below sets the number of bits in word used by MSBSP (serial port) for 
	receives from AIC23 (audio port). We are using a 32 bit packet containing two 
	16 bit numbers hence 32BIT is set for  receive */
	MCBSP_FSETS(RCR1, RWDLEN1, 32BIT);	

	/* Configures interrupt to activate on each consecutive available 32 bits 
	from Audio port hence an interrupt is generated for each L & R sample pair */	
	MCBSP_FSETS(SPCR1, RINTM, FRM);

	/* These commands do the same thing as above but applied to data transfers to the 
	audio port */
	MCBSP_FSETS(XCR1, XWDLEN1, 32BIT);	
	MCBSP_FSETS(SPCR1, XINTM, FRM);	
	

}
/********************************** init_HWI() **************************************/ 
void init_HWI(void)
{
	IRQ_globalDisable();			// Globally disables interrupts
	IRQ_nmiEnable();				// Enables the NMI interrupt (used by the debugger)
	IRQ_map(IRQ_EVT_RINT1,4);		// Maps an event to a physical interrupt
	IRQ_enable(IRQ_EVT_RINT1);		// Enables the event
	IRQ_globalEnable();				// Globally enables interrupts

}
        
/******************************** process_frame() ***********************************/  
void process_frame(void)
{
	int k, m; 
	int io_ptr0;
	
	//If low-pass magnitude enhancement is selected
	#if LOW_PASS_MAGNITUDE == 1
		//Defining parameter used for low-pass filtering mag(X)
		float kFrame;
	#endif 
	
	//If low-pass noise magnitude enhancement is selected
	#if LOW_PASS_NOISE == 1 
		//Defining parameter used for low-pass filtering mmse (i.e. mag(N))
		float kFrameNoise;
	#endif
	
	#if DELTA_TERM == 1
		//Define the float to hold the delta term
		float delta;
	#endif
	
	//Number of frames in m1 (i.e. the current MMSE bucket)
	static int countMin = 0;
	
	//Temporary pointer to swap minimum frame
	float *temp;

	//G(w) function used in noise subtraction part
	float gw;
	
	//If dynamic alpha enhancement is selected
	#if USE_DYNAMIC_ALPHA
		//Variable to store SNR for varible alpha
		float snr;
	#endif
	
	//Minimum mag(Y_Prev, Y_Curr, Y_Next)
	float minMag;
	
	//If low-pass magnitude enhancement is selected
	#if LOW_PASS_MAGNITUDE == 1
		//Assigning value to parameter used for low-pass filtering mag(X)
		//N.B. Use expf that calculates float instead of double to make
		//computation faster
		kFrame = expf(-TFRAME/tau);
	#endif
	
	//If low-pass noise magnitude enhancement is selected
	#if LOW_PASS_NOISE == 1
		//Assigning value to parameter used for low-pass filtering mmse
		//N.B. Use expf that calculates float instead of double to make
		//computation faster
		kFrameNoise = expf(-TFRAME/tauNoise);
	#endif
	
	/* work out fraction of available CPU time used by algorithm */    
	cpufrac = ((float) (io_ptr & (FRAMEINC - 1)))/FRAMEINC;  
		
	/* wait until io_ptr is at the start of the current frame */ 	
	while((io_ptr/FRAMEINC) != frame_ptr); 
	
	/* then increment the framecount (wrapping if required) */ 
	if (++frame_ptr >= (CIRCBUF/FRAMEINC)) frame_ptr=0;
 	
 	/* save a pointer to the position in the I/O buffers (inbuffer/outbuffer) where the 
 	data should be read (inbuffer) and saved (outbuffer) for the purpose of processing */
 	io_ptr0=frame_ptr * FRAMEINC;
	
	/* copy input data from inbuffer into inframe (starting from the pointer position) */ 
	 
	m=io_ptr0;
    for (k=0;k<FFTLEN;k++)
	{                           
		inframe[k] = inbuffer[m] * inwin[k]; 
		if (++m >= CIRCBUF) m=0; /* wrap if required */
	} 
	
	/************************* DO PROCESSING OF FRAME  HERE **************************/
	
	//Copy the contents of inframe to intermediate
	//Convert to complex for fft function
	for(k=0; k<FFTLEN; k++)
		intermediate[k] = cmplx(inframe[k],0);
	
	//Calculate the FFT of the current frame
	fft(FFTLEN, intermediate);
	
	//N.B. most of the frame processing is done withing a for loop
	//that takes advantage of the conjugate complex symmetry.
	//This greatly improves efficiency.
	for(k=0; k<FFTLEN/2; k++){
		//Calculate magnitude of current frame
		mag[k] = cabs(intermediate[k]);
		
		//If low-pass magnitude enhancement is selected
		#if LOW_PASS_MAGNITUDE == 1
			#if USE_MAGNITUDE_SQUARED == 1
				//N.B. Use powf that calculates float instead of double to make
				//computation faster
				lpfMagSquared[k] = (1-kFrame)*powf(mag[k],2) + kFrame*lpfMagSquared[k];
				//Save the low-passed filtered magnitude in a different array which is needed later
				lpfMag[k] = sqrtf(lpfMagSquared[k]);
			#else
				//Compute low-pass filtered magnitude
				lpfMag[k] = (1-kFrame)*mag[k] + kFrame*lpfMag[k];
			#endif
		#endif
		
		//If low-pass magnitude enhancement is selected
		#if LOW_PASS_MAGNITUDE == 1
			//Check for possible MMSE elements in current frame
			m1[k] = min(lpfMag[k],m1[k]);
		#else
			//Check for possible MMSE elements in current frame
			m1[k] = min(mag[k],m1[k]);
		#endif
		
		#if LOW_PASS_NOISE == 1
			//Calculate the low-pass filetered estimate of mmse (i.e. mag(N))
			mmse[k] = (1-kFrameNoise)*min(min(m1[k],m2[k]),min(m3[k],m4[k])) + kFrameNoise*mmse[k];
		#else
			//Calculate MMSE from MMSE buckets
			mmse[k] = min(min(m1[k],m2[k]),min(m3[k],m4[k]));
		#endif
		
		#if USE_DYNAMIC_ALPHA == 1
			//Calculate SNR
			snr = 20*log10f(lpfMag[k]/mmse[k]);
			
			//Implement piecewise scalling for alpha
			alpha = a0 - snr*s;
			
			//Check if alpha has exceeded the predefined limits
			if(alpha > alphaMax)
				alpha = alphaMax;
			else if(alpha < alphaMin)
				alpha = alphaMin;
		#endif
		
		#if DELTA_TERM == 1
			//Compute the delta term based on the piecewise function
			if(k < 32) delta = 1;
			else if(k < 64) delta = 2.5;
			else delta = 1.5;
		#elif DELTA_TERM == 0
			delta = 1;
		#endif
		
		//Choose different G(w); the different number associated with G_OMEGA are related to the order they 
		//are presented in the report with 0 being the first and 5 the last
		#if G_OMEGA == 0
			gw = max(lambda,(1-(delta*alpha*mmse[k]/mag[k])));
		#elif G_OMEGA == 1
			gw = max(lambda*(mmse[k]/mag[k]),(1-(delta*alpha*mmse[k]/mag[k])));
		#elif G_OMEGA == 2
			#if LOW_PASS_MAGNITUDE == 1
				gw = max(lambda*(lpfMag[k]/mag[k]),(1-(delta*alpha*mmse[k]/mag[k])));
			#else
				#error LOW_PASS_MAGNITUDE must be 1 when G_OMEGA = 2,3,4
			#endif
		//For values of G_OMEGA >= 3 some prerequisits apply. If these have not been
		//selected by the user, the compilation will fail and an approriate error is 
		//presented. This is used to improve robustness.
		#elif G_OMEGA == 3
			//If low-pass magnitude enhancement is selected
			#if LOW_PASS_MAGNITUDE == 1
				gw = max(lambda*(mmse[k]/lpfMag[k]),(1-(delta*alpha*mmse[k]/lpfMag[k])));
			#else
				#error LOW_PASS_MAGNITUDE must be 1 when G_OMEGA = 2,3,4
			#endif
		#elif G_OMEGA == 4
			//If low-pass magnitude enhancement is selected
			#if LOW_PASS_MAGNITUDE == 1
				gw = max((lambda),(1-(delta*alpha*mmse[k]/lpfMag[k])));
			#else
				#error LOW_PASS_MAGNITUDE must be 1 when G_OMEGA = 2,3,4
			#endif
		#elif G_OMEGA == 5
			//If low-pass magnitude enhancement is selected
			#if LOW_PASS_MAGNITUDE == 1 && USE_MAGNITUDE_SQUARED == 1
				gw = max((lambda),(sqrtf(1-(delta*alpha*powf(mmse[k],2)/lpfMagSquared[k]))));
			#else
				#error When G_OMEGA == 5, LOW_PASS_MAGNITUDE AND USE_MAGNITUDE_SQUARED must be 1
			#endif
		#else
			#error G_OMEGA must be: 0 <= G_OMEGA <= 5
		#endif
		
		//If residual noise enhancement is selected
		#if RESIDUAL_NOISE_REDUCTION == 1
			//Find the minimum magnitude of the previous, current and future magnitude spectrum
			minMag = min(min(cabs(Y_Prev[k]),cabs(Y_Curr[k])),cabs(rmul(gw,intermediate[k])));
			
			if(minMag == cabs(Y_Prev[k])){
				intermediate[k] = Y_Prev[k];
				intermediate[FFTLEN-1-k] = Y_Prev[FFTLEN-1-k];
			}else if(minMag == cabs(Y_Curr[k])){
				intermediate[k] = Y_Curr[k];
				intermediate[FFTLEN-1-k] = Y_Curr[FFTLEN-1-k];
			}else{
				intermediate[k] = rmul(gw,intermediate[k]);
				intermediate[FFTLEN-1-k] = rmul(gw,intermediate[FFTLEN-1-k]);
			}
			//Shift frames. Y_Next -> Y_Curr (gw) -> Y_Prev
			Y_Prev[k] = Y_Curr[k];
			Y_Prev[FFTLEN-1-k] = Y_Curr[FFTLEN-k-1];
			Y_Curr[k] = rmul(gw,intermediate[k]);
			Y_Curr[k] = rmul(gw,intermediate[FFTLEN-k-1]);
			
		#else
			intermediate[k] = rmul(gw,intermediate[k]);
			intermediate[FFTLEN-1-k] = rmul(gw,intermediate[FFTLEN-1-k]);
		#endif
		
	}
	
	//Check if current MMSE bucket if full
	if(++countMin >= BUCKET_FRAMES){
		countMin = 0; //Reset the counter
		
		//Reset oldest MMSE bucket
		for(k=0; k<FFTLEN/2; k++)
			m4[k] = mag[k];
		
		//Swap buckets
		temp = m4;
		m4 = m3;
		m3 = m2;
		m2 = m1;
		m1 = temp;
	}

	//Calculate the IFFT of the current frame
	ifft(FFTLEN, intermediate);
	
	//Copy real part of X[k] to the output frame
	for(k=0; k<FFTLEN; k++)
		outframe[k] = intermediate[k].r;
							      										
	/********************************************************************************/
	
    /* multiply outframe by output window and overlap-add into output buffer */  
                           
	m=io_ptr0;
    
    for (k=0;k<(FFTLEN-FRAMEINC);k++) 
	{    										/* this loop adds into outbuffer */                       
	  	outbuffer[m] = outbuffer[m]+outframe[k]*outwin[k];   
		if (++m >= CIRCBUF) m=0; /* wrap if required */
	}         
    for (;k<FFTLEN;k++) 
	{                           
		outbuffer[m] = outframe[k]*outwin[k];   /* this loop over-writes outbuffer */        
	    m++;
	}	                                   
}        
/*************************** INTERRUPT SERVICE ROUTINE  *****************************/

// Map this to the appropriate interrupt in the CDB file
   
void ISR_AIC(void)
{       
	short sample;
	/* Read and write the ADC and DAC using inbuffer and outbuffer */
	
	sample = mono_read_16Bit();
	inbuffer[io_ptr] = ((float)sample)*ingain;
		/* write new output data */
	mono_write_16Bit((int)(outbuffer[io_ptr]*outgain)); 
	
	/* update io_ptr and check for buffer wraparound */    
	
	if (++io_ptr >= CIRCBUF) io_ptr=0;
}

/************************************************************************************/

float max(const float a, const float b){
	return (a<b)?b:a;
}

float min(const float a, const float b){
	return (a>b)?b:a;
}
\end{lstlisting}
\end{Huge}
%Replace listings caption name (i.e. Listing) with Appendix
\renewcommand\lstlistingname{Appendix}
\setcounter{figure}{1}
\end{document}